\documentclass[aps,prl,twocolumn,superscriptaddress,nofootinbib]{revtex4-2}

\usepackage{latexsym}
\usepackage{amsmath}
\usepackage{amssymb}
\usepackage{amsfonts}
\usepackage{bm}
\usepackage{appendix}
\RequirePackage{lineno}

\usepackage{color}
\definecolor{purple}{rgb}{0.5,0,0.5}
\definecolor{blue}{rgb}{0.0,0,0.9}
\definecolor{prdblue}{rgb}{0.133,0.118,0.498}
\usepackage[colorlinks=true, pdfstartview=FitV, linkcolor=prdblue, citecolor= prdblue, urlcolor=prdblue]{hyperref}

\usepackage{supertabular}
\usepackage{placeins}
\usepackage{epsfig}
\usepackage{graphicx}
\usepackage{hyperref}
\usepackage{comment}
\usepackage{physics}
\usepackage{multirow}
\usepackage{float}
\usepackage{dcolumn}
\usepackage{siunitx}
\usepackage{subfigure}
\usepackage{makecell}

\hypersetup{
  breaklinks=true,
  colorlinks = true,
  linkcolor = blue,
  anchorcolor = blue,
  citecolor = blue,
  filecolor = blue,
  pagecolor = blue,
  urlcolor = blue
}
\hypersetup{
  bookmarks=true,
  bookmarksnumbered=true,
  bookmarkstype=toc,
  linktocpage=true
}

\newcommand{\LumiPsip            }{2712.4 \pm 14.3} 
\newcommand{\BRLambdaPpi          }{63.90 \pm 0.50}  
\newcommand{\NumPsip              }{2712.4 \pm 14.3}  

\newcommand{\BRSWave}{2.11 \pm 0.08 ^{+0.07}_{-0.18}}
\newcommand{\BRPWave}{0.81 \pm 0.08 ^{+0.07}_{-0.20}}

\newcommand{\BWPMass }{2262 \pm 34}
\newcommand{\BWPWidth}{215  \pm 35}

\newcommand{\EventsPsipA    }{107.7 \pm 0.6}
\newcommand{\EventsPsipB    }{345.4 \pm 2.6}
\newcommand{\EventsPsipC    }{2259.3 \pm 11.1}

\def \romanOne   {\uppercase\expandafter{\romannumeral1}}
\def \romanTwo   {\uppercase\expandafter{\romannumeral2}}
\def \romanThree {\uppercase\expandafter{\romannumeral3}}
\def \romanFour  {\uppercase\expandafter{\romannumeral4}}
\def \romanFive  {\uppercase\expandafter{\romannumeral5}}
\def \romanSix   {\uppercase\expandafter{\romannumeral6}}
\def \romanSeven {\uppercase\expandafter{\romannumeral7}}
\def \romanEight {\uppercase\expandafter{\romannumeral8}}

\begin{document}


\setlength{\oddsidemargin}{-0.5cm} \addtolength{\topmargin}{15mm}

\title{Observation of $\eta(2600)$ and Threshold Enhancements in the $\Lambda\bar{\Lambda}$ System}
\newcommand{\BESIIIorcid}[1]{\href{https://orcid.org/#1}{\hspace*{0.1em}\raisebox{-0.45ex}{\includegraphics[width=1em]{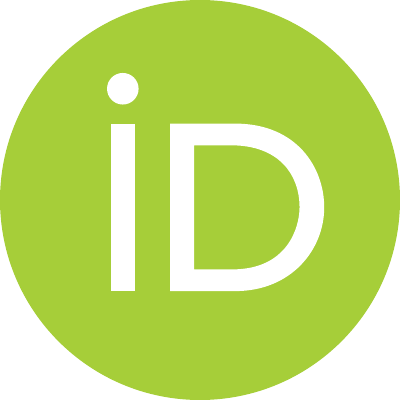}}}}

\author{
  \begin{small}
    \begin{center}
M.~Ablikim$^{1}$\BESIIIorcid{0000-0002-3935-619X},
M.~N.~Achasov$^{4,b}$\BESIIIorcid{0000-0002-9400-8622},
P.~Adlarson$^{81}$\BESIIIorcid{0000-0001-6280-3851},
X.~C.~Ai$^{87}$\BESIIIorcid{0000-0003-3856-2415},
C.~S.~Akondi$^{31A,31B}$\BESIIIorcid{0000-0001-6303-5217},
R.~Aliberti$^{39}$\BESIIIorcid{0000-0003-3500-4012},
A.~Amoroso$^{80A,80C}$\BESIIIorcid{0000-0002-3095-8610},
Q.~An$^{77,64,\dagger}$,
Y.~H.~An$^{87}$\BESIIIorcid{0009-0008-3419-0849},
Y.~Bai$^{62}$\BESIIIorcid{0000-0001-6593-5665}
% ...
\\
\vspace{0.2cm}
(BESIII Collaboration)\\
\vspace{0.2cm} {\it
$^{1}$ Institute of High Energy Physics, Beijing 100049, People's Republic of China\\
% ...
$^{4}$ Budker Institute of Nuclear Physics SB RAS (BINP), Novosibirsk 630090, Russia\\
% ...
$^{31}$ INFN Sezione di Ferrara, (A)INFN Sezione di Ferrara, I-44122, Ferrara, Italy; (B)University of Ferrara, I-44122, Ferrara, Italy\\
% ...
$^{39}$ Johannes Gutenberg University of Mainz, Johann-Joachim-Becher-Weg 45, D-55099 Mainz, Germany\\
% ...
$^{62}$ Southeast University, Nanjing 211100, People's Republic of China\\
% ...
$^{64}$ State Key Laboratory of Particle Detection and Electronics, Beijing 100049, Hefei 230026, People's Republic of China\\
% ...
$^{77}$ University of Science and Technology of China, Hefei 230026, People's Republic of China\\
% ...
$^{80}$ University of Turin and INFN, (A)University of Turin, I-10125, Turin, Italy; (B)University of Eastern Piedmont, I-15121, Alessandria, Italy; (C)INFN, I-10125, Turin, Italy\\
$^{81}$ Uppsala University, Box 516, SE-75120 Uppsala, Sweden\\
% ...
$^{87}$ Zhengzhou University, Zhengzhou 450001, People's Republic of China\\

\vspace{0.2cm}
$^{\dagger}$ Deceased\\
% ...
$^{b}$ Also at the Novosibirsk State University, Novosibirsk, 630090, Russia\\
% ...

}

\end{center}
    \vspace{0.4cm}
\end{small}}

\begin{abstract}
Using $\LumiPsip$ million $\psi(3686)$ events collected with the BESIII detector, the $\Lambda\bar{\Lambda}$ system produced in $\psi(3686)$ radiative decays is studied. A model-independent partial wave analysis reveals a significant threshold enhancement structure dominated by the $^1S_0$ and $^3P_0$ partial waves, corresponding to $J^{PC} = 0^{-+}$ and $0^{++}$, respectively. In addition, a new pseudoscalar resonance, designated as $\eta(2600)$, is observed in the $^1S_0$ partial wave with a mass value consistent with the previously reported $X(2600)$ state, which represents the heaviest light meson observed to date. These results enhance our understanding of baryon-antibaryon threshold dynamics and the pseudoscalar light hadron spectroscopy.
\end{abstract}

\maketitle

The study of threshold enhancements in baryon-antibaryon systems has attracted considerable attention over the past decades. This phenomenon was first observed in 2003 by the BES Collaboration in $J/\psi \to \gamma p\bar{p}$~\cite{BES:2003aic}, and subsequently confirmed by the CLEO and BESIII experiments with improved statistics~\cite{CLEO:2010fre,BESIII:2010vwa, BESIII:2011aa}. The properties of the $p\bar{p}$ threshold enhancement depend on its production mechanism: in the reactions $e^+e^- \to p\bar{p}$~\cite{BESIII:2019hdp} and $e^+e^- \to p\bar{p}\pi^0$~\cite{BESIII:2017qwj}, the $p\bar{p}$ system has negative C-parity; whereas in the decay process $J/\psi \to \gamma p\bar{p}$ ~\cite{BES:2003aic}, it has positive C-parity. Experimentally, the data are described by introducing near-threshold resonances with different masses and spin-parities, such as $\rho^{*}(p\bar{p})$ ~\cite{BESIII:2017qwj} and $X(p\bar{p})$ ~\cite{BESIII:2011aa}. Final-state interactions have also been used to describe several $p\bar{p}$ threshold structures~\cite{Salnikov:2023ipo}.

The near-threshold enhancement in $\Lambda\bar{\Lambda}$ systems was first observed by the Belle Collaboration in the decays $B \to \Lambda\bar{\Lambda}K^{(*)0}$~\cite{Belle:2008kpj}. Subsequent experimental investigations have been conducted mainly by BESIII through measurements of $e^+e^- \to \Lambda\bar{\Lambda}$~\cite{BESIII:2017hyw}, $e^+e^- \to \Lambda\bar{\Lambda}\phi$~\cite{BESIII:2021fqx}, and $e^+e^- \to \Lambda\bar{\Lambda}\eta$~\cite{BESIII:2022tvj}. Similar to the $p\bar{p}$ case, these $\Lambda\bar{\Lambda}$ threshold enhancements appear in various quantum number configurations. The measured lineshapes are well reproduced using a $\Lambda\bar{\Lambda}$ final-state interaction model in Ref.~\cite{Haidenbauer:2023zcu}. A predicted $\Lambda\bar{\Lambda}$ baryonium~\cite{Wan:2021vny} is also a possible explanation of those phenomena.

Determining the spin-parity quantum numbers of baryon-antibaryon systems, particularly distinguishing $0^{-+}$ from $0^{++}$ configurations, requires measurement of spin correlations between the baryon and antibaryon. For spin-0 baryon-antibaryon final states, the angular distribution remains insensitive to parity variations, and parity information is encoded solely in their spin correlations. Experimental extraction of these spin correlations is challenging in $p\bar{p}$ systems, due to lack of proton spin polarimeter. In contrast, the weak decays $\Lambda \to p\pi^{-}$ and $\bar{\Lambda} \to \bar{p}\pi^{+}$ act as effective spin analyzers, enabling direct measurement of spin correlations in $\Lambda\bar{\Lambda}$ systems through their self-analyzing property. This capability allows clear differentiation between $\Lambda\bar{\Lambda}$ threshold enhancements with distinct $J^{PC}$ quantum numbers.

The $\Lambda\bar{\Lambda}$ system also provides an excellent probe for investigating the pseudoscalar meson spectrum above the $\Lambda\bar{\Lambda}$ threshold, as baryon-antibaryon systems produced via charmonium radiative decays are dominated by  the $^1S_0$ partial wave with $J^{PC} = 0^{-+}$. Several pseudoscalar candidates have been observed in this region, including the $\eta(2225)$~\cite{BESIII:2016qzq}, $\eta(2370)$~\cite{BESIII:2023wfi}, and $X(2600)$~\cite{BESIIICollaboration:2022kwh}. The $X(2600)$, observed so far only in the $J/\psi \to \gamma\pi^+\pi^-\eta'$ channel, requires confirmation in additional decay modes to establish its properties and determine its spin-parity quantum numbers.

The nature of the $X(2600)$ has inspired diverse theoretical interpretations. It has been proposed as a $D$-wave strangeonium state~\cite{Lodha:2024qby}, while other exotic explanations include $ss\bar{q}\bar{q}$ or $sq\bar{s}\bar{q}$ tetraquark configurations~\cite{Lodha:2024yfn}, and a $2^{-+}$ $us\bar{u}\bar{s}$ tetraquark state~\cite{Wang:2025nme}. More notably, some studies suggest a predominantly gluonic nature, identifying the $X(2600)$ as a candidate for a $2^{-+}$ glueball~\cite{Zhang:2022obn} or a $0^{-+}$ glueball~\cite{Giacosa:2023fdz}. Precise determination of the spin-parity and searching for additional decay modes are therefore essential to elucidate its internal structure.

In this work, we perform a model-independent partial wave analysis (PWA) of the decay $\psi(3686) \to \gamma \Lambda\bar{\Lambda}$ using $(\NumPsip) \times 10^6$ $\psi(3686)$ events collected with the BESIII detector, aiming to determine the partial wave structure of the $\Lambda\bar{\Lambda}$ system near the production threshold.

Detailed descriptions of the BESIII detector and BEPCII collider are provided in Refs.~\cite{BESIII:2009fln,Yu:2016cof,BESIII:2020nme,Huang:2022wuo}. This analysis employs three $\psi(3686)$ data samples collected in 2009, 2012, and 2021, with $(\EventsPsipA)$, $(\EventsPsipB)$, and $(\EventsPsipC)$ million events, respectively~\cite{BESIII:2024lks}, together with their corresponding inclusive Monte Carlo (MC) samples. For the decay $\psi(3686) \to \gamma\Lambda\bar{\Lambda}$, a signal MC sample of 10 million events is generated based on a PWA model, which performs a global fit to the data using a coherent sum of Breit--Wigner (BW) resonances~\cite{supple}.
Final-state radiation (FSR) from charged particles is incorporated using the {\textsc{PHOTOS}} package~\cite{photos}.

The signal process follows the decay chain $\psi(3686) \to \gamma X$, $X \to \Lambda\bar{\Lambda}$, with $\Lambda \to p\pi^-$ and $\bar{\Lambda} \to \bar{p}\pi^+$, where $X$ denotes an intermediate state. This sequence yields four charged tracks reconstructed in the multilayer drift chamber (MDC) and one isolated photon shower detected in the electromagnetic calorimeter (EMC). Well-reconstructed tracks in the MDC must satisfy $\vert\!\cos\theta\vert < 0.93$, where $\theta$ is the polar angle relative to the beam axis. The distance of closest approach along the beam direction between each track and the interaction point is required to be less than 20 cm. These selection criteria ensure that the tracks lie within the detector’s acceptance region and originate from the primary collision vertex. The number of good charged tracks is required to be at least four.

Particle identification is performed by combining the specific ionization  energy loss (${\rm d}E/{\rm d}x$) measured in the MDC and the flight time from the time-of-flight (TOF) system to compute likelihoods $\mathcal{L}(h)$ for each hadron hypothesis $h$ ($h = p, \pi$)~\cite{Asner:2008nq}. A track is identified as a proton if $\mathcal{L}(p) > \mathcal{L}(\pi)$ and $\mathcal{L}(p) > 0$; otherwise, it is classified as a pion.

The $p$, $\bar{p}$, $\pi^+$, and $\pi^-$ tracks are uniquely assigned by minimizing $\sqrt{(M_{\Lambda} - m_{\Lambda})^2 + (M_{\bar{\Lambda}} - m_{\bar{\Lambda}})^2}$, where $M_{\Lambda}$ ($M_{\bar{\Lambda}}$) denotes the invariant mass of the $p\pi^{-}$ ($\bar{p}\pi^+$) pair and $m_{\Lambda}$ ($m_{\bar{\Lambda}}$) represents the known $\Lambda$ ($\bar{\Lambda}$) mass~\cite{pdg2024}. Vertex fits are applied to both $p\pi^-$ and $\bar{p}\pi^+$ combinations with a requirement of $\chi^2 < 100$. Furthermore, secondary vertex fits~\cite{Xu:2009zzg} constrain the $\Lambda$ momentum direction to align with the vector from the decay vertex to the interaction point. To suppress non-$\Lambda$ backgrounds, the $\Lambda$ decay length divided by its uncertainty is required to be greater than 2.0. Identical selection criteria are applied to $\bar{\Lambda}$ candidates.

Photon showers in the EMC are considered valid if the deposited energy exceeds 25 MeV in the barrel region ($\vert\!\cos\theta\vert < 0.8$) or 50 MeV in the endcap region ($0.86 < \vert\!\cos\theta\vert < 0.92$). The time difference between the EMC shower and the event start time is required to be within [0, 700] ns. Each qualified shower is combined with the four  charged tracks for a four-constraint (4C) kinematic fit (KF), enforcing total four-momentum conservation according to the initial $\psi(3686)$. The shower yielding the least $\chi^2_{4C}$ from the KF is selected as the radiative photon candidate, and $\chi^2_{4C} < 40$ is required.

To suppress background contributions from multi-photon processes, an alternative 4C KF combining with two photons is performed. All possible combinations of the four selected charged tracks with two showers are tested, and the minimum  $\chi^2$ is obtained. Events are rejected if $\chi^2$ of the two-photon 4C KF is less than that of the single-photon case, as they likely originate from processes with two photons. However, studies based on the inclusive MC samples indicate that certain multi-photon processes still contribute to background events, primarily from $\psi(3686) \to \pi^0\Lambda\bar{\Lambda}$, $ \gamma\Lambda\bar{\Sigma}^0$, and $ \gamma\bar{\Lambda}\Sigma^0$. These background channels are thoroughly investigated based on control samples selected in data, as detailed  in the Supplemental material~\cite{supple}.

\begin{figure}[tbp]
  \centering
  \subfigure
  {
      \includegraphics[width=0.47\textwidth]{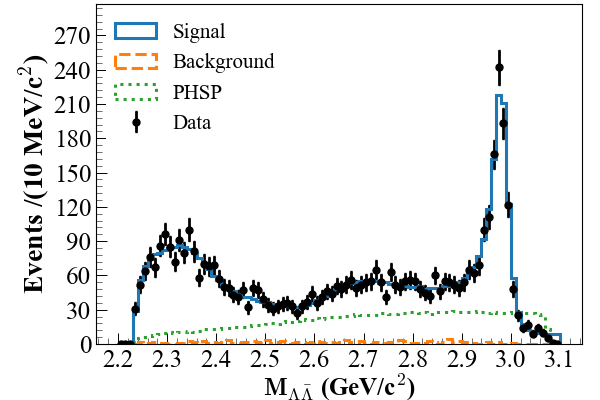}
  }
  \caption{Distribution of the $\Lambda\bar{\Lambda}$ invariant mass, including combined backgrounds from the $\pi^0\Lambda\bar{\Lambda}$, $\gamma\Lambda\bar{\Sigma}^0$, and $\gamma\bar{\Lambda}\Sigma^0$ processes in the inclusive MC samples. The ``signal'' denotes signal MC events generated based on the PWA model.}
  \label{fig:invariant_mass_low}
\end{figure}

The $M_{\Lambda\bar{\Lambda}}$ distributions for data, signal MC, phase space (PHSP) MC, and background MC are presented in Fig.~\ref{fig:invariant_mass_low}. This analysis is restricted to the near-threshold region $M_{\Lambda\bar{\Lambda}} \in [2.231, 3.080]$~GeV/$c^2$, and the data sample in this region contains 4896 events with an estimated background contribution of 117 events, corresponding to a signal purity of 97.6\%.

A model-independent PWA is performed to study the $\Lambda\bar{\Lambda}$ system. The $\Lambda\bar{\Lambda}$ invariant mass spectrum is partitioned into 40 bins, consisting of 30 equal-statistics bins between 2.230 and 2.940 GeV/$c^2$, 9 equal-interval bins spanning (2.940, 3.046) GeV/$c^2$, and one dedicated bin for the (3.046, 3.080) GeV/$c^2$ region. Finer binning is applied around the $\eta_c$ region to resolve its narrow peak. An independent partial wave fit is conducted in each bin.

The construction of partial wave amplitudes follows the formalism described in Refs.~\cite{BESIII:2022udq,BESIII:2024mbf}. For a two-body decay process $0 \to 1 + 2$, the helicity amplitudes~\cite{Chung:1997jn} are expressed as
\begin{equation}
    A_{\lambda_0\lambda_1\lambda_2}(\phi,\theta) = H_{\lambda_1\lambda_2} D_{\lambda_0,\lambda_1-\lambda_2} ^{J_0*}(\phi,\theta,0),
\label{equ:helicityamp}
\end{equation}
where  $\phi$ and  $\theta$ are the helicity angles defined in the helicity frame of particle 0; $\lambda_0$, $\lambda_1$, and $\lambda_2$ are the helicities of particles 0, 1, and 2, respectively. $D$ is the Wigner-$D$ function. The helicity amplitude $H_{\lambda_1,\lambda_2}$ is expanded to  the LS coupling formula~\cite{Chung:1997jn} as

\begin{equation}
\small
\begin{aligned}
    & H_{\lambda_1 \lambda_2}  = \sum_{ls} g_{ls}\sqrt{\frac{2l+1}{2J_0+1}} \langle l0,s\delta|J_0\delta\rangle \langle J_1 \lambda_1,J_2-\lambda_2|s\delta \rangle ,
    \label{s04:ls_coupling}
\end{aligned}
\end{equation}
where $g_{ls}$ represents the LS coupling amplitudes for the partial wave with total spin $s$ and orbital angular momentum $l$, $J_{0,1,2}$ denote the spins of particles 0, 1, and 2, respectively. $\delta=\lambda_1 - \lambda_2$ is the helicity difference between particles 1 and 2.

The angular-dependent amplitude for the decay chain $\psi(3686) \to \gamma X$, $X \to \Lambda\bar{\Lambda}$, with $\Lambda \to p\pi^-$ and $\bar{\Lambda} \to \bar{p}\pi^+$, can be expressed as
\begin{equation}
\small
\begin{aligned}
A^{X}_{\lambda_{\rm ext}} (\Omega)  = \sum_{\lambda_{X}\lambda_{\Lambda}\lambda_{\bar{\Lambda}}} A_{\lambda_{\psi(3686)}\lambda_\gamma\lambda_X} (\phi_\gamma,\theta_\gamma) A_{\lambda_X\lambda_{\Lambda}\lambda_{\bar{\Lambda}}} (\phi_\Lambda, \theta_\Lambda ) \\
 \times A_{\lambda_{\Lambda}\lambda_{p}} (\phi_p, \theta_p) A_{\lambda_{\bar{\Lambda}}\lambda_{\bar{p}}} (\phi_{\bar{p}}, \theta_{\bar{p}}),
\end{aligned}
\end{equation}
where $\lambda_{\rm ext} = ( \lambda_{\psi(3686)},\lambda_\gamma,\lambda_{p}, \lambda_{\bar{p}}$) denotes the helicities of the external (initial and final) particles, $\Omega$ = $(\phi_\gamma,\theta_\gamma, \phi_\Lambda, \theta_\Lambda, \phi_p, \theta_p, \phi_{\bar{p}}, \theta_{\bar{p}})$ denotes all helicity angles in the decay chain. The amplitudes $A_{\lambda_{\Lambda}\lambda_{p}}$ and $A_{\lambda_{\bar{\Lambda}}\lambda_{\bar{p}}}$ are constrained using the $\Lambda$ and $\bar{\Lambda}$ decay parameters measured by BESIII~\cite{BESIII:2022qax}. Further details are provided in the Supplemental material~\cite{supple}.

The energy-dependent component of the amplitude for the $j$-th partial wave in the $i$-th bin is described by a complex parameter $c_{ij} = \rho_{ij} e^{i\alpha_{ij}}$, where $\rho_{ij}$ is the magnitude and $\alpha_{ij}$ is the phase. Then the total amplitude in the $i$-th bin can be expressed as
\begin{equation}
    A_{\lambda_{\rm ext},i} (\Omega) = \sum_{j} c_{ij}  A^{X_j}_{\lambda_{\rm ext}} (\Omega),
\label{equ:total_amplitude}
\end{equation}

Averaging over the initial-state helicities and summing over the final-state helicities, the probability function is given by

\begin{equation}
    |A|^2(\Omega)  = \frac{1}{2}\sum_{\lambda_{\rm ext}}  |A_{\lambda_{\rm ext}}(\Omega)|^2.
\label{equ:total_amplitude_square}
\end{equation}
When only the $^1S_0$ and $^3P_0$ partial waves are included, their amplitudes are strictly orthogonal, resulting in the absence of interference terms in the total amplitude. This orthogonality renders the phase parameters ${\alpha_{ij}}$ redundant; thus, they are fixed to zero in the fit, leaving $\rho_{ij}$ as the only free parameters.

The goal of this analysis is to determine $\rho_{ij}$ for each partial wave in each bin by fitting the data. The amplitude squared, $\rho_{ij}^2$, which directly represents the intensity of each partial wave, will be reported. The distribution of $\rho_{ij}^2$ characterizes the amplitude structure of each partial wave in the $\Lambda\bar{\Lambda}$ system in a model-independent manner.

The extended maximum likelihood method~\cite{Barlow:1990vc} is used to estimate the $\rho_{ij}$, with the likelihood construction following Ref.~\cite{BESIII:2022udq}. The fitting procedure is carried out using the open-source software TF-PWA (TensorFlow-based PWA software)~\cite{Jiang:2024vbw}.

The extended negative log-likelihood is given by
\begin{equation}
\begin{split}
    \mathcal{S} &=  \int |A|^2(\Omega) d\Phi+\ln N! - \sum s(\Omega) \ln |A|^2(\Omega),
\label{equ:finalNLL}
\end{split}
\end{equation}
where $N$ is the net number of signal events in data, and the first term involves the integration of the squared amplitude modulus over the reconstructed PHSP $\Phi$, which is computed using detector-simulated MC events. The last term represents a summation over the combined data and background samples. The weight function $s(\Omega)$ takes the value 1 for data events, $-0.08$ for $\pi^0\Lambda\bar{\Lambda}$ background, $-0.17$ for $\gamma\Lambda\bar{\Sigma}^0$ background, $-0.29$ for $\gamma\bar{\Lambda}\Sigma^0$ background, and $-1$ for other inclusive background components. These weight factors are implemented to correct the branching fractions (BFs) of the corresponding background processes in the inclusive MC samples, with further details provided in the Supplemental material~\cite{supple}. Continuum background is neglected in the fitting procedure.

\begin{figure}[tbp]
    \centering
    \includegraphics[width=0.45\textwidth]{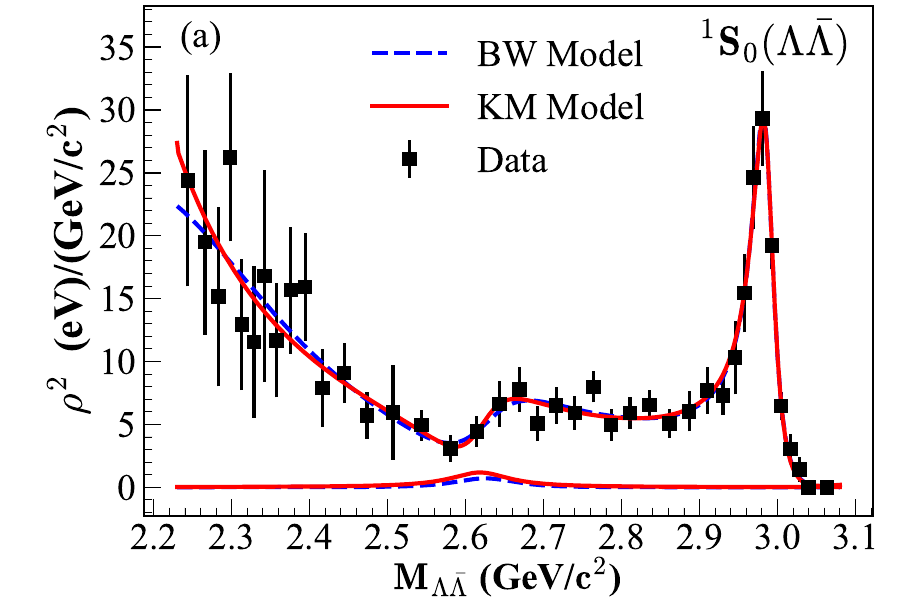}
    \includegraphics[width=0.45\textwidth]{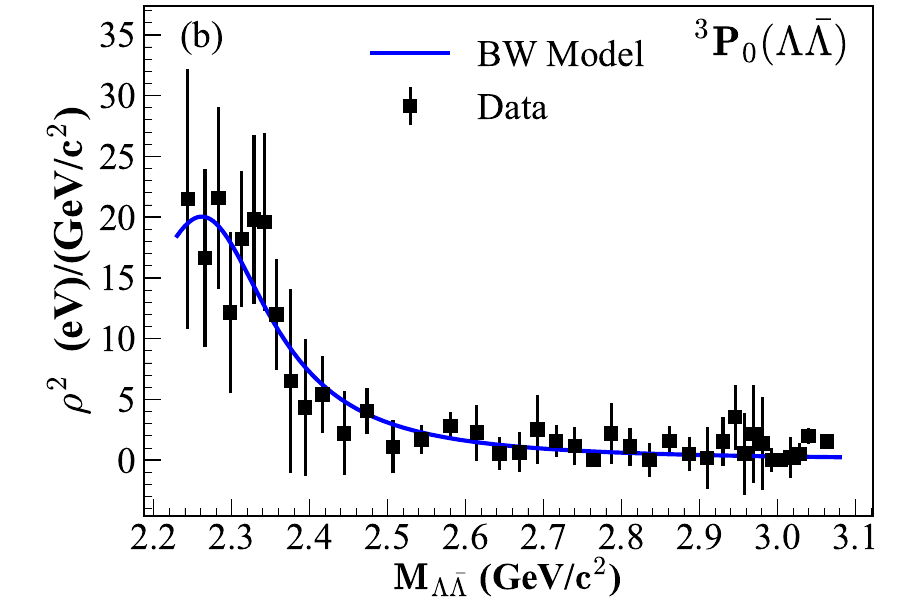}
    \caption{The values and uncertainties of $\rho^2$ for the (a) $^1S_0$ and (b) $^3P_0$ partial waves, along with fit results. For the $^1S_0$ partial wave, both the BW model (blue dashed) and KM model (red solid) are shown. For the $^3P_0$ partial wave, a single BW fit (blue solid) is performed. The isolated resonance component near 2.6~GeV/$c^2$ in the $^1S_0$ partial wave corresponds to the $\eta(2600)$ resonance.}
    \label{fig:model_fits}
\end{figure}

To determine which partial waves should be included in the fit, we test five partial waves: $^1S_0$, $^3P_0$, $^3P_1$, $^3P_2$, and $^1D_2$. The statistical significance of each partial wave is assessed by comparing the change in the log-likelihood and the degrees of freedom between fits with all five waves included and those with each wave individually removed. Then the significances are computed via the likelihood ratio test method~\cite{Wilks:1938dza}, and they are determined to be 13.2$\sigma$, 5.4$\sigma$, 0.1$\sigma$, 2.9$\sigma$, and 1.1$\sigma$, respectively. The $^1S_0$ and $^3P_0$ waves show significances above 5$\sigma$, the $^3P_2$ wave has a significance slightly below 3$\sigma$, while the $^3P_1$ and $^1D_2$ waves are statistically insignificant. Consequently, the final PWA includes only the $^1S_0$ and $^3P_0$ waves. The measured values of ${\rho_{ij}^2}$ and their total uncertainties for these two waves are shown in Fig.~\ref{fig:model_fits}.

Systematic uncertainties are evaluated from multiple sources. The uncertainty from additional partial waves is estimated by performing alternative fits including the $^3P_2$ wave. The assumption of constant energy-dependent amplitudes within each bin is tested by introducing a linear variation model, $\rho_{ij}^{2\prime} = \rho_{ij}^{2} + k_{ij}(m - m_i)$, where $m_i$ is the center mass of the $i$-th bin and $k_{ij}$ is a free parameter. Background-related uncertainties are assessed by both removing the background component and by incorporating continuum background estimated from 20.3~fb$^{-1}$ of data collected at $\sqrt{s} = 3.773$~GeV. The uncertainties due to detector efficiency are evaluated using $J/\psi \to \mu\mu\gamma_{\mathrm{ISR}}$ initial-state radiation (ISR) control samples for photon efficiency corrections and $\psi(3686) \to \Lambda\bar{\Lambda}$ control samples for $\Lambda$ and $\bar{\Lambda}$ reconstruction efficiency corrections. For each source, the largest deviation between the alternative and nominal results is assigned as the systematic uncertainty.

The product BF for the $j$-th partial wave is defined by $\mathcal{B}_j = \left( \sum_{i} N_{i,j}^{\text{truth}} \right) / \left( N_{\psi(3686)} \cdot \mathcal{B}_{\Lambda}^2 \right)$, where $N_{\psi(3686)} = (\NumPsip) \times 10^6$~\cite{BESIII:2024lks} represents the total number of $\psi(3686)$ events, and $\mathcal{B}_{\Lambda} = (\BRLambdaPpi)\%$~\cite{pdg2024} is the BF of $\Lambda \to p\pi^-$. Here, $N_{ij}^{\text{truth}} = N_{\text{signal},ij} / \epsilon_i$ corresponds to the efficiency-corrected signal yield, with $\epsilon_i$ denoting the efficiency of the $i$-th bin, calculated from signal MC samples generated using the PWA model.

The systematic uncertainties on the BF from various sources are summarized in Table~\ref{tab:systematics}. The product BFs are determined to be $\mathcal{B}(^1S_0)=(\BRSWave) \times 10^{-5}$ for the $^1S_0$ partial wave and $\mathcal{B}(^3P_0)=(\BRPWave) \times 10^{-5}$ for the $^3P_0$ partial wave, where the first and second uncertainties are statistical and systematic, respectively. The dominant contribution to the systematic uncertainty originates from the inclusion of additional partial waves, contributing $-$7.8\% for $^1S_0$ and $-$22.8\% for $^3P_0$.

\begin{table}[htbp]
\setlength{\tabcolsep}{11.5pt}
\centering
\caption{Relative systematic uncertainties (in \%) on the product BFs $\mathcal{B}[\psi(3686) \to \gamma X, X \to \Lambda\bar{\Lambda}]$ for $^1S_0$ and $^3P_0$ partial waves.}
\label{tab:systematics}
\begin{tabular}{lcc}
\hline\hline
Source & $^1S_0$ (\%) & $^3P_0$ (\%) \\
\hline
Additional partial waves & $-$7.8 & $-$22.8 \\
Total $\psi(3686)$ events & 0.6 & 0.6 \\
$\mathcal{B}(\Lambda \to p\pi^-)$ & 1.6 & 1.6 \\
Constant assumption of $\rho_{ij}$ & 2.1 & 5.4 \\
Non-physical background & 1.8 & 4.5 \\
Efficiency correction & 0.1 & 0.2 \\
QED background & 0.8 & 3.2 \\
\hline
\rule{0pt}{3ex}Total systematic uncertainty & $^{+3.3}_{-8.5}$ & $^{+7.8}_{-24.1}$ \\[2pt]
\hline\hline
\end{tabular}
\end{table}

The $\rho_{ij}^2$ values in each bin, with total uncertainties combining statistical and systematic components, are displayed as black error bars in Fig.~\ref{fig:model_fits}. The $^1S_0$ partial wave accurately reproduces the $\eta_c$ peak, which is consistent with the $J^{PC} = 0^{-+}$ quantum numbers and confirms the validity of our analysis method.
We report the normalized $\rho^2_j$ distribution subject to the constraint $\int \rho^2_{j}(m) f(m) \mathrm{d}m = \Gamma_{\psi(3686)} \mathcal{B}_j$, where $\Gamma_{\psi(3686)} = 286~\text{keV}$ is the total width of the $\psi(3686)$ and $f(m)$ is the PHSP factor. We take $f(m) = p(m) q(m) / (p_0 q_0)$, where $p(m)$ and $q(m)$ are the momenta of the $\Lambda\bar{\Lambda}$ system and the $\Lambda$ baryon, respectively, evaluated in the rest frame of their parent particle, and $p_0$ and $q_0$ denote their values at $m = 2.6~\text{GeV}/c^2$. The $\rho^2_j$ distribution directly represents the intrinsic lineshape of the partial wave amplitudes, independent of detector efficiency and kinematic PHSP constraints.

To describe the lineshapes and extract resonance parameters, a fit is performed to the measured $\rho^2_{ij}$ values taking into account their statistical and systematic uncertainties. Two distinct dynamical models are implemented. The BW model employs a superposition of three BW propagators. The $K$-matrix (KM) model uses a single-channel three-pole KM parametrization. Further details of these formulations are provided in the Supplemental material~\cite{supple}.

\begin{figure}[tbp]
    \centering
    \includegraphics[width=0.47\textwidth]{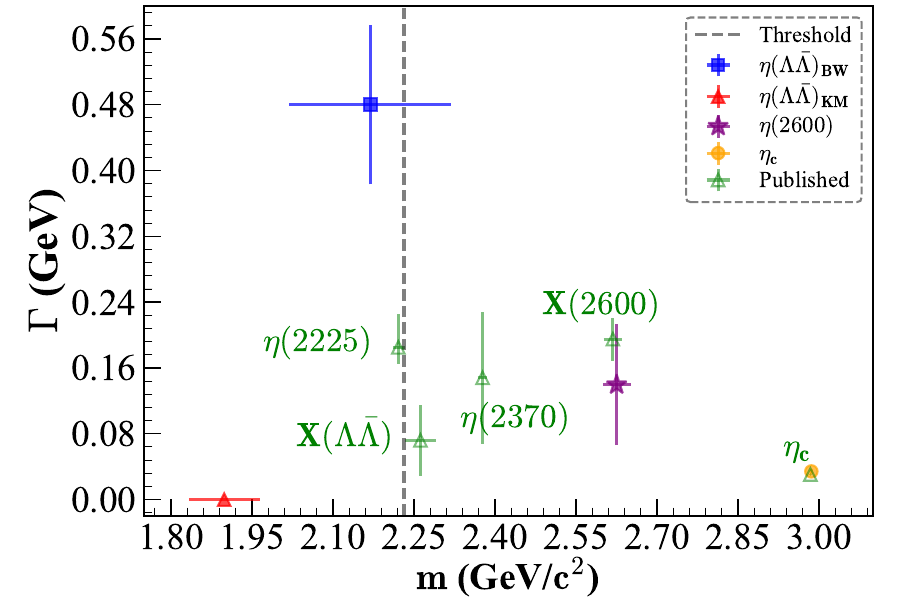}
    \caption{The resonance masses and widths of the $^1S_0$ partial wave, as obtained from two different models, compared with previously published results ($X(\Lambda\bar{\Lambda})$ from Ref.~\cite{BESIII:2021fqx}, $\eta(2225)$, $\eta(2370)$ from Ref.~\cite{pdg2024} and $X(2600)$ from Ref.~\cite{BESIIICollaboration:2022kwh}). For $\eta(\Lambda\bar{\Lambda})$: blue squares represent the BW model; red triangles represent the KM model. Purple stars and orange circles represent the $\eta(2600)$ and $\eta_c$, respectively. Green open triangles: previously reported states. The gray dashed line represents the $\Lambda\bar{\Lambda}$ threshold.}
    \label{fig:poles}
\end{figure}

The fitting results are presented in Fig.~\ref{fig:model_fits}(a). Both models provide a good description of the data with comparable fit quality. Both the total fit and the resonance components near 2.6~GeV/$c^2$ are shown in the figure. The extracted resonance masses and widths are shown in Fig.~\ref{fig:poles} and listed in Table~\ref{tab:poles}, alongside several previously reported states, including the $X(\Lambda\bar{\Lambda})$ state observed by BESIII~\cite{BESIII:2021fqx}, $\eta(2225)$, $\eta(2370)$, $X(2600)$, and $\eta_c$. The product BFs for $\psi(3686) \to \gamma X_{k} \to \gamma\Lambda\bar{\Lambda}$ are also determined and presented in Table~\ref{tab:poles}, with the calculation methodology detailed in the Supplemental material~\cite{supple}. For the BW model, the masses and widths are obtained directly as the fit parameters. For the KM model, the pole positions are first extracted numerically from the fit results and subsequently converted to BW masses and widths using the relation $s_{\text{pole}} = m^2 - i m \Gamma$.

A clear threshold enhancement structure is observed, which we refer to as $\eta(\Lambda\bar{\Lambda})$. This feature can be described either by a broad BW resonance or, alternatively, using a subthreshold pole. Such a threshold enhancement is of particular interest as it may be interpreted within the framework of hexaquark states or baryonium configurations, with theoretical predictions suggesting a pseudoscalar $\Lambda\bar{\Lambda}$ bound state around $(2.27 \pm 0.13)$~GeV/$c^2$~\cite{Wan:2021vny}. Furthermore, we identify a resonance near 2.6~GeV/$c^2$, labeled $\eta(2600)$, with a measured mass of $2625 \pm 26 \pm 6$~MeV/$c^2$ and width of $140 \pm 58 \pm 46$~MeV, where the first uncertainties correspond to the fitted uncertainties and the second are due to the choice of the BW and KM lineshape models. The mass is consistent with that of $X(2600)$~\cite{BESIIICollaboration:2022kwh}, which represents the heaviest light meson observed to date~\cite{pdg2024}, within uncertainty. The statistical significance of the $\eta(2600)$ is 7.0$\sigma$, as evaluated by a model-dependent PWA fit to the full data set \cite{supple}. Notably, the product BF $\mathcal{B}(\psi(3686) \to \gamma \eta(2600) \to \gamma\Lambda\bar{\Lambda})$ lies in the range $(0.56\text{--}1.01) \times 10^{-6}$ depending on the model, representing $(12\text{--}22)\%$ of that for $\eta_c$ in the same final state. This ratio is comparable to that of $X(2600)$ relative to $\eta_c$ in $J/\psi \to \gamma\pi^+\pi^-\eta'$, which is $(31.0\,^{+8.7}_{-9.8})\%$ as calculated from the measured BFs in Ref.~\cite{BESIIICollaboration:2022kwh}. This consistency in the relative production rates provides additional evidence supporting the interpretation that $\eta(2600)$ and $X(2600)$ are the same resonance.

\begin{table}
\centering
\setlength{\tabcolsep}{9pt}
\caption{The measured masses (in MeV$/c^2$) and widths (in MeV) of the resonances obtained from the model-independent PWA. Here, $\mathcal{B}_{k}$ denotes the product BF of $\psi(3686) \to \gamma X_{k}$ with $X_{k} \to \Lambda\bar{\Lambda}$, given in units of $10^{-6}$. The $\chi^2/\mathrm{ndf}$ values indicate the fit quality for each model, where ndf is the number of degrees of freedom.}
\label{tab:poles}
\begin{tabular}{lccl}
\hline
\hline
 & BW Model & KM Model & Unit \\
\hline
$\chi^2/\mathrm{ndf}$ & 15.4/29 & 15.8/30 & - \\
$m_{\eta(\Lambda\bar{\Lambda})}$ & 2169 $\pm$ 150 & 1899 $\pm$ 66 & MeV/$c^2$ \\
$\Gamma_{\eta(\Lambda\bar{\Lambda})}$ & 480 $\pm$ 97 & - & MeV \\
$m_{\eta(2600)}$ & 2625 $\pm$ 26 & 2620 $\pm$ 17 & MeV/$c^2$ \\
$\Gamma_{\eta(2600)}$ & 140 $\pm$ 58 & 94 $\pm$ 36 & MeV \\
$m_{\eta_c}$ & 2985.8 $\pm$ 1.4 & 2986.2 $\pm$ 1.3 & MeV/$c^2$ \\
$\Gamma_{\eta_c}$ & 34.2 $\pm$ 2.8 & 34.3 $\pm$ 2.7 & MeV \\
$\mathcal{B}_{\eta(\Lambda\bar{\Lambda})}$ & $14.5^{+0.8}_{-1.4}$ & $14.2^{+0.8}_{-1.4}$ & $10^{-6}$ \\
$\mathcal{B}_{\eta(2600)}$ & $1.01^{+0.05}_{-0.09}$ & $0.56^{+0.03}_{-0.05}$ & $10^{-6}$ \\
$\mathcal{B}_{\eta_c}$ & $4.66^{+0.24}_{-0.44}$ & $4.51^{+0.23}_{-0.42}$ & $10^{-6}$ \\
\hline
\hline
\end{tabular}
\end{table}

For the $^3P_0$ partial wave fit, the overall lineshape demonstrates only one prominent threshold enhancement structure. Therefore, we consider only a single BW resonance. The fitting results are shown in panel (b) of Fig.~\ref{fig:model_fits}. The BW model gives a mass of $\BWPMass~\text{MeV}/c^2$ and a width of $\BWPWidth~\text{MeV}$. This $^3P_0$ state lies close to the $f_0(2200)$~\cite{pdg2024}, which has a mass of $2187\pm14~\text{MeV}/c^2$ and a width of $210\pm40~\text{MeV}$.

In summary, we have performed a model-independent PWA of the $\Lambda\bar{\Lambda}$ system produced in $\psi(3686)$ radiative decays. The system is dominated by the $^1S_0$ and $^3P_0$ partial waves, both of which exhibit significant threshold enhancements. The squared amplitudes of both partial waves are reported as a function of $M_{\Lambda\bar{\Lambda}}$.
In the $^1S_0$ partial wave ($J^{PC} = 0^{-+}$), we observe a prominent threshold enhancement structure, $\eta(\Lambda\bar{\Lambda})$, as well as the $\eta(2600)$ resonance. The measured mass of the $\eta(2600)$ is $2625 \pm 26 \pm 6$~MeV/$c^2$ and the width is $140 \pm 58 \pm 46$~MeV, where the first uncertainties are from the fit and the second are from the choice of lineshape model. Both the mass and width are consistent within uncertainties with the previously observed $X(2600)$ in $J/\psi \to \gamma\pi^+\pi^-\eta'$~\cite{BESIIICollaboration:2022kwh}. The comparable relative production rates of the $\eta(2600)$ and $X(2600)$ with respect to $\eta_c$ in their respective channels further support the interpretation that they are the same resonance. Our results lend support to theoretical interpretations suggesting the existence of pseudoscalar glueball-like states around 2.6~GeV~\cite{Giacosa:2023fdz}. The significant threshold enhancements observed in both the $^1S_0$ and $^3P_0$ partial waves provide valuable insights into near-threshold baryon-antibaryon dynamics.

\vspace{0.2cm}

The BESIII Collaboration thanks the staff of BEPCII (https://cstr.cn/31109.02.BEPC) and the IHEP computing center for their strong support. This work is supported in part by National Key R\&D Program of China under Contracts Nos. 2025YFA1613900, 2023YFA1606000, 2023YFA1606704; National Natural Science Foundation of China (NSFC) under Contracts Nos. 12575112, 11635010, 11935015, 11935016, 11935018, 12025502, 12035009, 12035013, 12061131003, 12192260, 12192261, 12192262, 12192263, 12192264, 12192265, 12221005, 12225509, 12235017, 12342502, 12361141819; the Chinese Academy of Sciences (CAS) Large-Scale Scientific Facility Program; the Strategic Priority Research Program of Chinese Academy of Sciences under Contract No. XDA0480600; CAS under Contract No. YSBR-101; 100 Talents Program of CAS; The Institute of Nuclear and Particle Physics (INPAC) and Shanghai Key Laboratory for Particle Physics and Cosmology; ERC under Contract No. 758462; German Research Foundation DFG under Contract No. FOR5327; Istituto Nazionale di Fisica Nucleare, Italy; Knut and Alice Wallenberg Foundation under Contracts Nos. 2021.0174, 2021.0299, 2023.0315; Ministry of Development of Turkey under Contract No. DPT2006K-120470; National Research Foundation of Korea under Contract No. NRF-2022R1A2C1092335; National Science and Technology fund of Mongolia; Polish National Science Centre under Contract No. 2024/53/B/ST2/00975; STFC (United Kingdom); Swedish Research Council under Contract No. 2019.04595; U. S. Department of Energy under Contract No. DE-FG02-05ER41374.

\bibliographystyle{apsrev4-2-noarxiv}
\bibliography{References}


\onecolumngrid

\clearpage
\newpage
\appendix

\textbf{\boldmath\large Supplemental material for ``Observation of $\eta(2600)$ and Threshold Enhancements in the $\Lambda\bar{\Lambda}$ System''}

\begin{appendices}

\section{Appendix A: Background Estimation}
\label{sec:bg_estimation}

The dominant backgrounds arise from the process $\psi(3686)\rightarrow\gamma\gamma\Lambda\bar{\Lambda}$, mainly through the channels $\psi(3686) \to \pi^0\Lambda\bar{\Lambda}$, $ \gamma\Lambda\bar{\Sigma}^0$, and $ \gamma\bar{\Lambda}\Sigma^0$. To better quantify their contributions, we employ a data-driven approach using these channels as control samples. The ratios of data to inclusive MC yields from these control samples are used to correct the normalization of the background MC samples, ensuring a reliable estimation of the total background yield in the signal channel. Table~\ref{tab:fit_bg} summarizes the values of $N_{\text{data}}$, $N_{\text{incl}}$, and the data-MC ratio for these three channels. The total background yield is estimated from the reweighted inclusive MC samples.

\begin{table}[H]
    \centering
    \small
    \caption{Background yields in the data and inclusive MC samples. $N_{\mathrm{data}}$ is the number of events in data, $N_{\mathrm{incl}}$ is the number of events in the inclusive MC samples, and $N_{\mathrm{data}}/N_{\mathrm{incl}}$ is the ratio of data to inclusive MC samples.}
    \label{tab:fit_bg}
    \begin{tabular}{cccc}
    \hline
    \hline
    Background Control Channel & $\pi^0\Lambda\bar{\Lambda}$ & $\gamma\Lambda\bar{\Sigma}^0$ & $\gamma\bar{\Lambda}\Sigma^0$ \\
    \hline
    $N_{\mathrm{data}}$    & 385 &  147 & 247 \\
    $N_{\mathrm{incl}}$      & 4804   &  853    & 840 \\
    $N_{\mathrm{data}}/N_{\mathrm{incl}}$ & 0.08   &  0.17   & 0.29 \\
    \hline
    \hline
    \end{tabular}
\end{table}

\section{Appendix B: Decay parameter of $\Lambda$ and $\bar{\Lambda}$}

The decay parameters of $\Lambda$ and  $\bar{\Lambda}$ are defined using the helicity amplitude $H_\lambda$ :
\begin{equation}
\begin{aligned}
    \alpha_{\Lambda}      &= \frac{|H^{\Lambda}_{+}|^2 - |H^{\Lambda}_{-}|^2  }{|H^{\Lambda}_{+}|^2 + |H^{\Lambda}_{-}|^2  }  , \\
    \alpha_{\bar\Lambda} &= \frac{|H^{\bar{\Lambda}}_{+}|^2 - |H^{\bar{\Lambda}}_{-}|^2  }{|H^{\bar{\Lambda}}_{+}|^2 + |H^{\bar{\Lambda}}_{-}|^2  } ,
\end{aligned}
\end{equation}
where the subscripts $+$ and $-$ denote the helicity $\lambda=\frac{1}{2}$ and $-\frac{1}{2}$ for proton and anti-proton, respectively. The helicity amplitude for the decays of $\Lambda$ and $\bar{\Lambda}$ can be expanded within the LS coupling scheme as follows:
\begin{equation}
\begin{aligned}
    H^{\Lambda}_{+} &= \frac{\sqrt{2}}{2} g_{(0,\frac{1}{2})} - \frac{\sqrt{2}}{2} g_{(1,\frac{1}{2})} , \\
    H^{\Lambda}_{-} &= \frac{\sqrt{2}}{2} g_{(0,\frac{1}{2})} + \frac{\sqrt{2}}{2} g_{(1,\frac{1}{2})} , \\
    H^{\bar{\Lambda}}_{+} &= \frac{\sqrt{2}}{2} \bar{g}_{(0,\frac{1}{2})} - \frac{\sqrt{2}}{2} \bar{g}_{(1,\frac{1}{2})} , \\
    H^{\bar{\Lambda}}_{-} &= \frac{\sqrt{2}}{2} \bar{g}_{(0,\frac{1}{2})} + \frac{\sqrt{2}}{2} \bar{g}_{(1,\frac{1}{2})} .
\end{aligned}
\end{equation}
To ensure consistency with the measured value $\alpha = -\bar{\alpha} = 0.7542$ reported by BESIII~\cite{BESIII:2022qax}, we fix the parameter $g_{(l,s)}$ as follows:
\begin{equation}
\begin{aligned}
    g_{(0,\frac{1}{2})}  &= 1.0 , \\
    g_{(1,\frac{1}{2})}  &= 0.4552575 , \\
    \bar{g}_{(0,\frac{1}{2})} &= 1.0 , \\
    \bar{g}_{(1,\frac{1}{2})} &= - 0.4552575 .\\
\end{aligned}
\end{equation}

\section{Appendix C: Likelihood Function}
The probability density function for observed events is given by:
\begin{equation}
    P (\Omega) =  \frac{|A|^2 (\Omega) \epsilon(\Omega) }{\int |A|^2 (\Omega) \epsilon(\Omega) d\Phi} ,
\label{equ:pdf}
\end{equation}
where $\Omega$ denotes the helicity angles, $\Phi$ is the standard PHSP for the $\psi(3686)\to \gamma \Lambda\bar\Lambda, \Lambda\to p\pi^-$ and $\bar\Lambda\to \bar  p\pi^+$ decays, and $\epsilon(\Omega)$ represents the efficiency function. The integration can be calculated by the MC method:
\begin{equation}
    \nu  = \int |A(\Omega)|^2 \epsilon(\Omega) d\Phi = \int |A(\Omega)|^2 d\Phi_\text{rec},
\label{equ:nu}
\end{equation}
where $\Phi_\text{rec}$ is the PHSP for the reconstructed events. The likelihood function is defined as the product of the probability density functions over all events:
\begin{equation}
    L = \prod_{k} P(\Omega_k)^{s(\Omega_k)},
\end{equation}
where $k$ denotes the event index in the combined data and background samples, and $s(\Omega_k)$ represents the weighting function, defined as:
\begin{equation}
    s(\Omega_k) = \begin{cases}
        1    &  \quad \Omega_k \in \text{data}, \\
        -0.08 &  \quad \Omega_k \in  \text{inclusive}\ \pi^0\Lambda\bar{\Lambda}\ \text{bg}, \\
        -0.17 &  \quad \Omega_k \in  \text{inclusive}\ \gamma\Lambda\bar{\Sigma}^0 \ \text{bg}, \\
        -0.29 &  \quad \Omega_k \in  \text{inclusive}\ \gamma\bar{\Lambda}\Sigma^0 \ \text{bg}, \\
        -1    &  \quad \Omega_k \in  \text{other inclusive bg}, \\
    \end{cases}
\label{equ:weight_function}
\end{equation}
where the negative weight factor for the background indicates that the background contribution is subtracted from the data. The weight factors for the $\pi^0\Lambda\bar{\Lambda}$, $\gamma\Lambda\bar{\Sigma}^0$, and $\gamma\bar{\Lambda}\Sigma^0$ channels are obtained through an analysis of the data and inclusive MC samples, as described in the Background Estimation section. The sum of $s(\Omega_k)$ should equal the number of signal events in the data:
\begin{equation}
    N = \sum_{k} s(\Omega_k) = N_\text{data} - N_\text{bg}.
\end{equation}
The extended likelihood function is defined as:

\begin{equation}
    \mathcal{L}^\text{extend} = P(\nu;N) \mathcal{L} = \frac{\nu^{N} e^{-\nu}}{N!}  \mathcal{L},
\label{equ:extended_likelihood}
\end{equation}
where $P(\nu;N)$ denotes the Poisson probability function. The extended likelihood function incorporates the constraint that the integral of the squared amplitude, $\nu$, follows a Poisson distribution with a mean value of $N$:

\begin{equation}
    \nu= \int |A(\Omega)|^2 d\Phi_\text{rec} \sim P(\nu;N).
\end{equation}
This constraint introduces an additional condition, allowing the integral of the squared amplitude to be interpreted physically as the ``number of signal events". Under this constraint, fixing the magnitude of the first partial-wave amplitude is no longer necessary; however, the phase remains redundant, necessitating that the phase of the first partial wave (i.e., the reference wave) is set to zero.

The extended negative log-likelihood function is given by:

\begin{equation}
\begin{split}
    \mathcal{S} &=-\ln \mathcal{L}^\text{extend} \\
    &= \nu  + \ln(N!)    -    \sum_{k} s(\Omega_k) \ln |A(\Omega_k)|^2,
\label{equ:extended_nll}
\end{split}
\end{equation}
where $N$ is independent of the fit parameters and can therefore be omitted during the fitting procedure.

\section{Appendix D: Model-Independent PWA Results}

\begin{figure}
    \centering
    \includegraphics[width=0.28\textwidth]{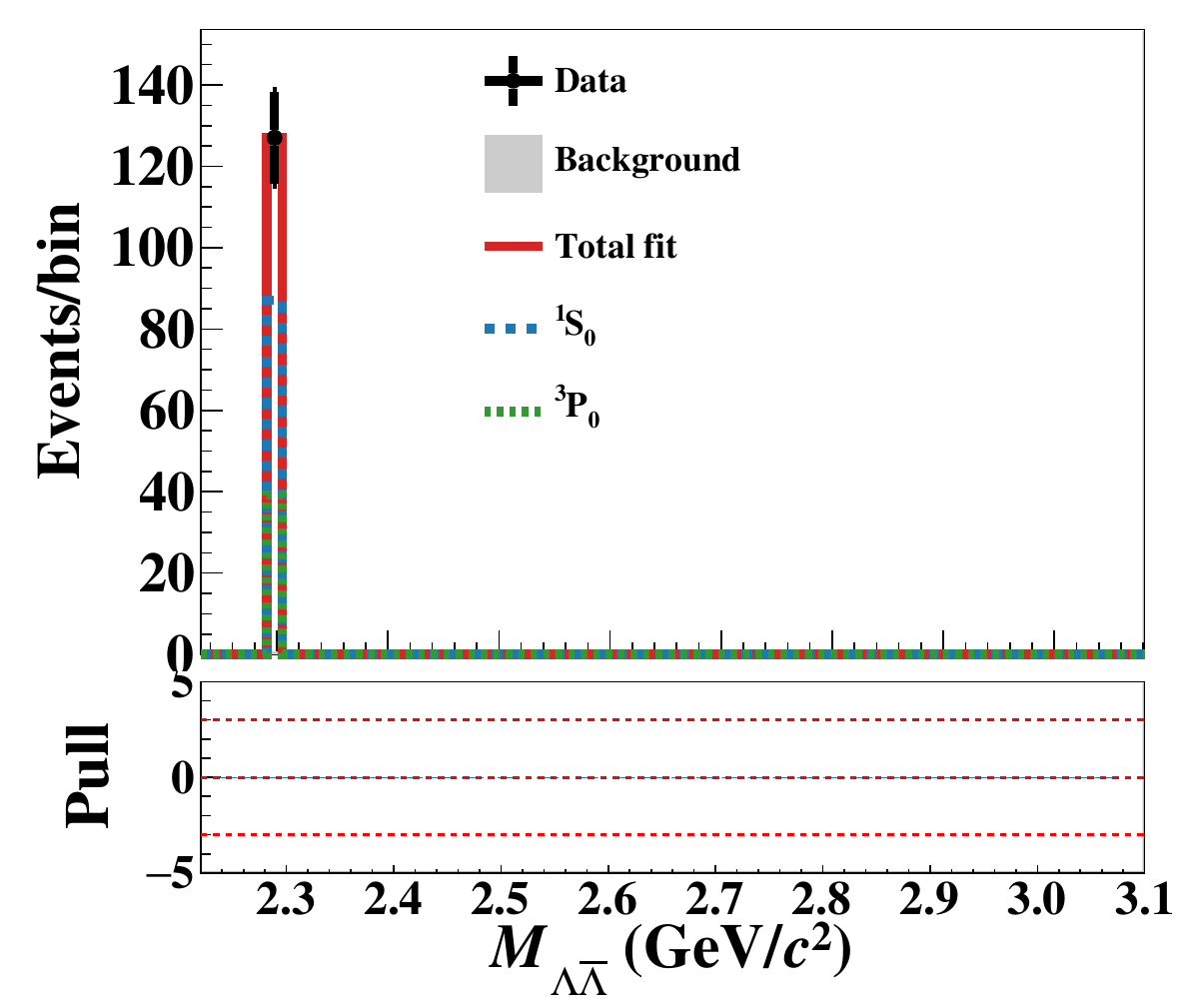}
    \includegraphics[width=0.28\textwidth]{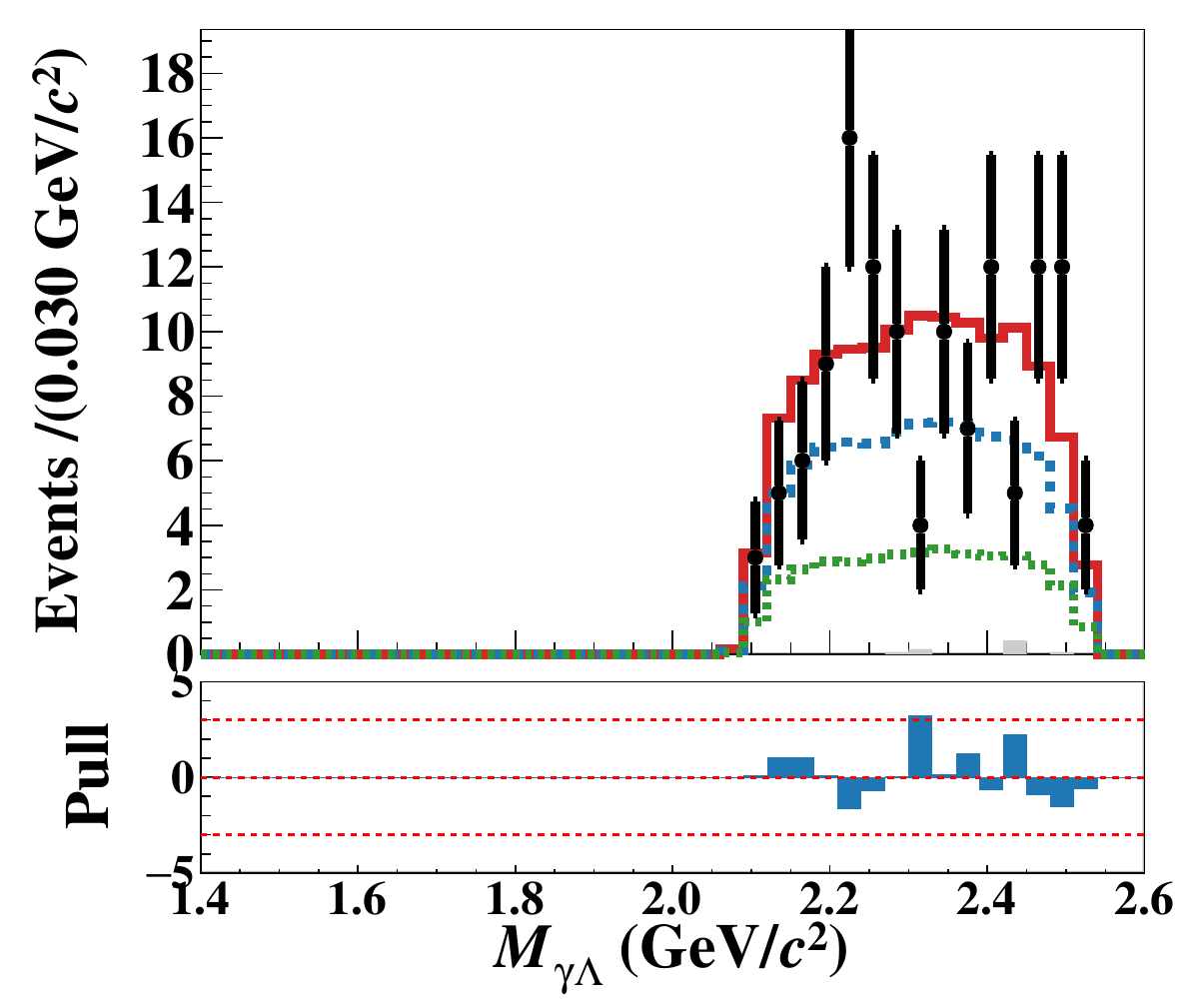}
    \includegraphics[width=0.28\textwidth]{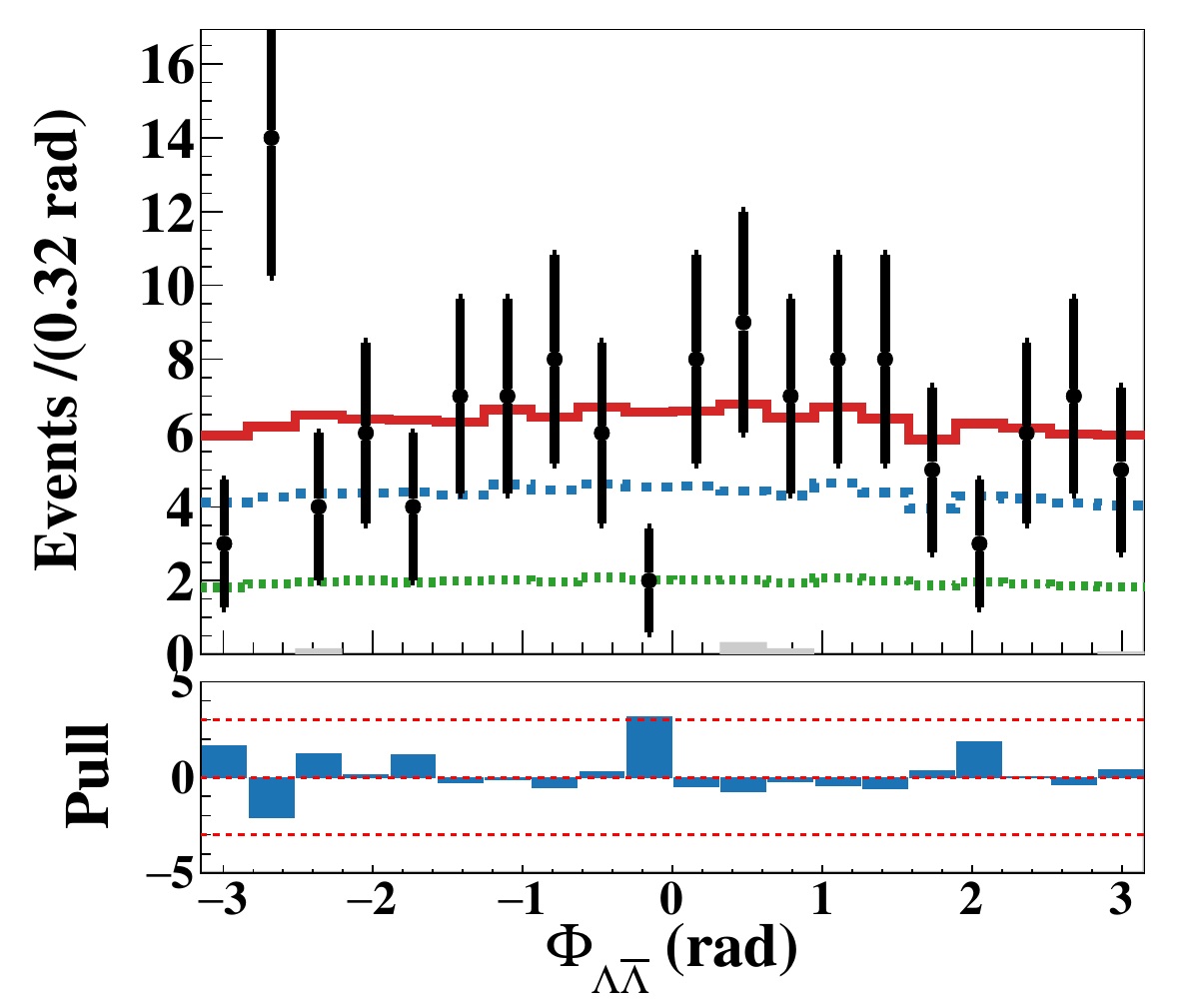}
    \includegraphics[width=0.28\textwidth]{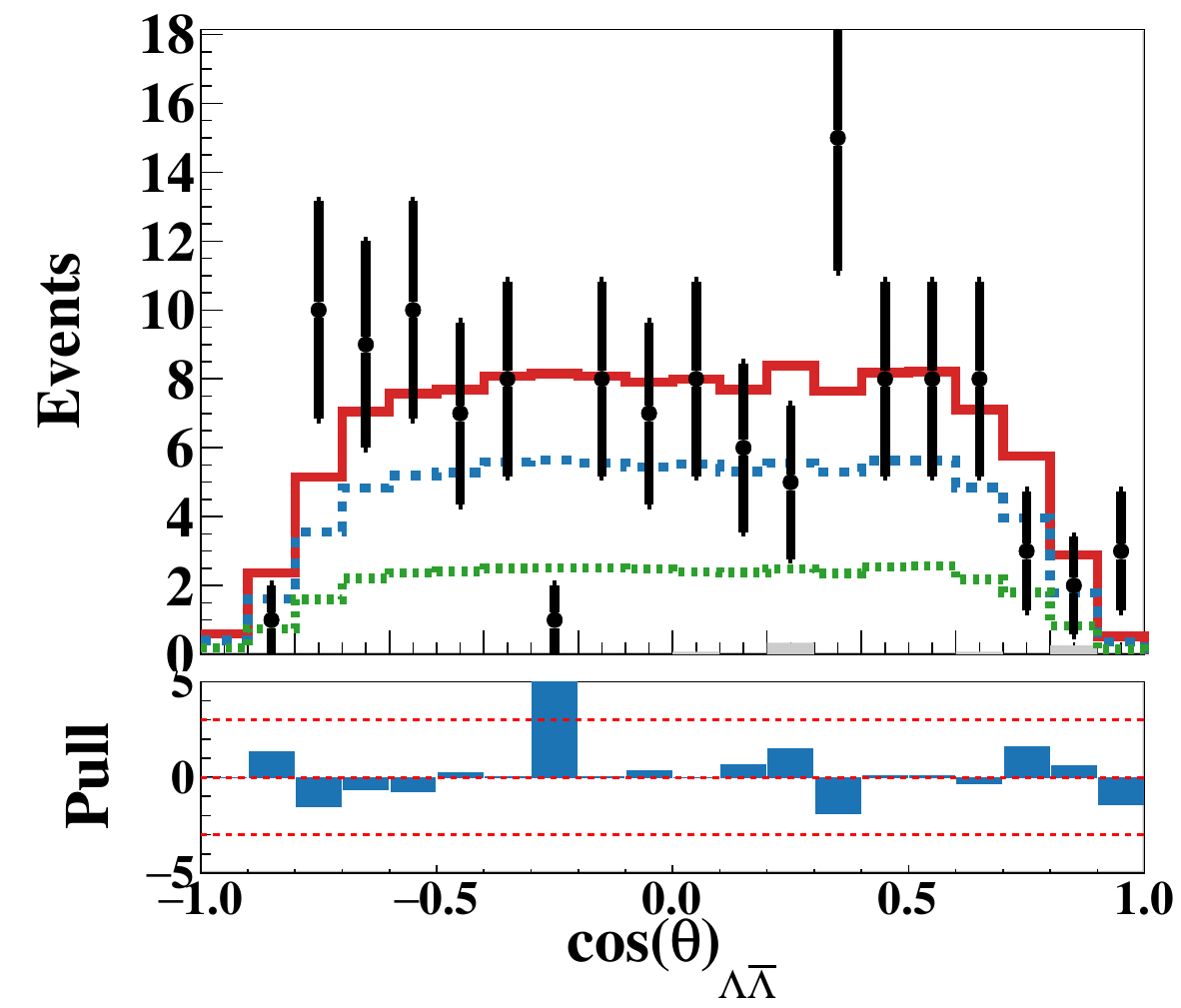}
    \includegraphics[width=0.28\textwidth]{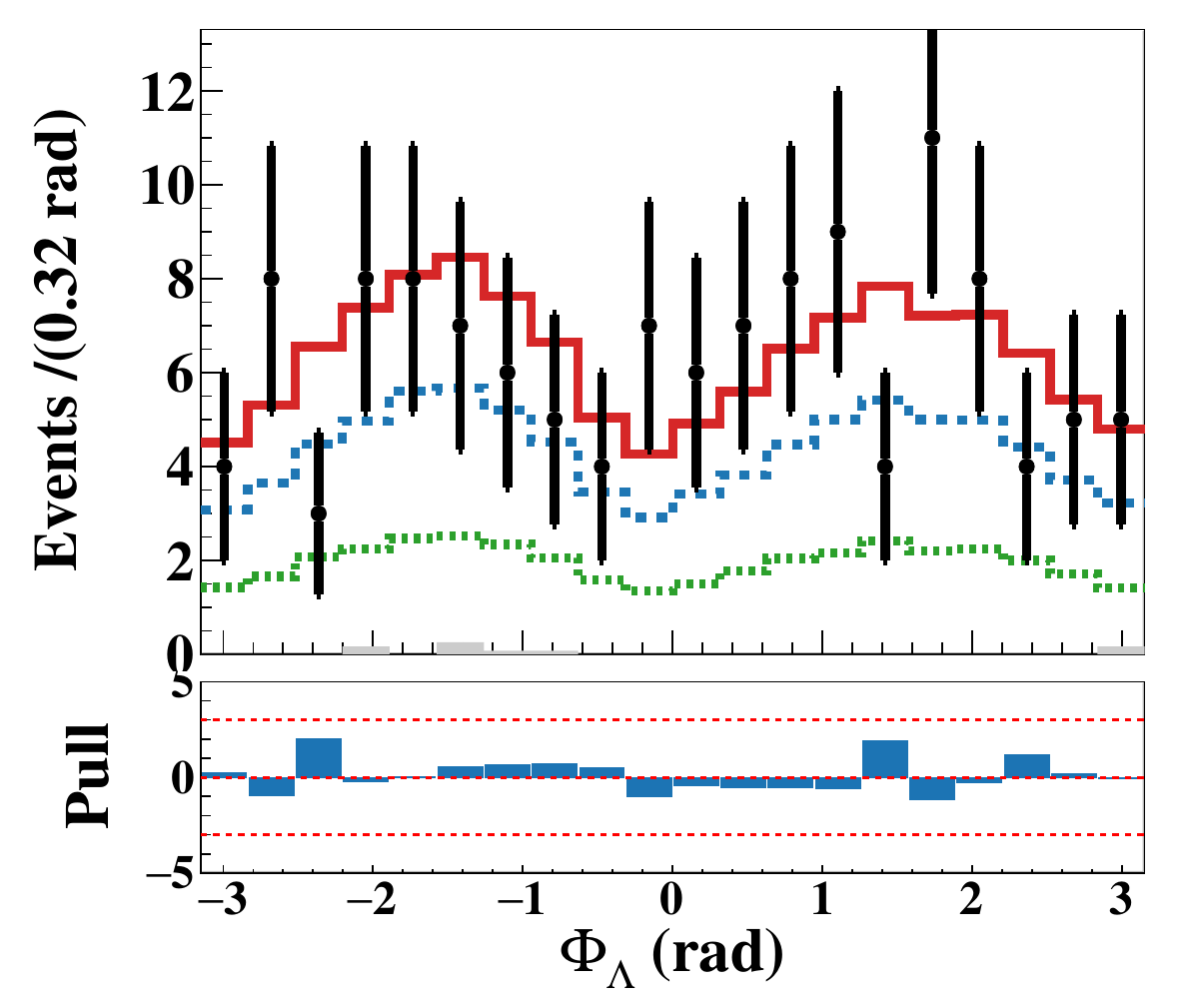}
    \includegraphics[width=0.28\textwidth]{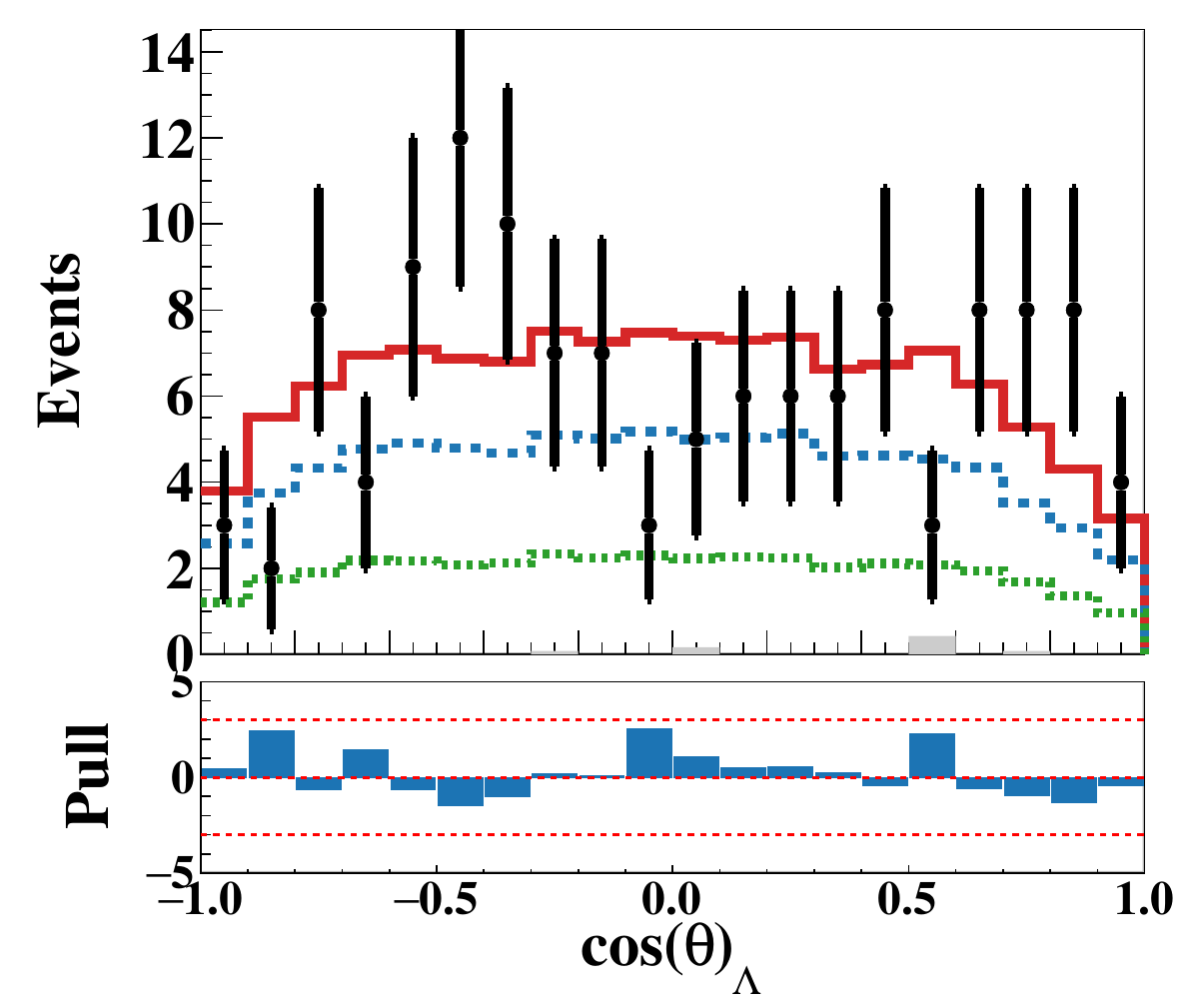}
    \caption{Fit results of the $0$-th bin of data.}
    \label{fig:fit_results}
\end{figure}

The fit results from the PWA for the $0$-th bin are presented in Fig.~\ref{fig:fit_results}.

Table~\ref{tab:amp_square} summarizes the bin range, number of data events, background events, signal events, efficiency, and the measured squared amplitude ($\rho^2$) for the $^1S_0$ and $^3P_0$ partial waves in each mass bin.

The values of $\rho^2_j$ are normalized to the partial width of $\psi(3686)\to \gamma X_j,\ X_j\to \Lambda\bar{\Lambda}$:
\begin{equation}
    \int \rho^2_{j}(m) f(m) \mathrm{d}m = \Gamma_{\psi(3686)} \mathcal{B}_j,
\end{equation}
where $f(m) = p(m) q(m) / (p_0 q_0)$, $p(m)$ and $q(m)$ are the momenta of the $\Lambda\bar{\Lambda}$ system and the $\Lambda$ baryon, respectively, evaluated in the rest frame of their parent particle, and $p_0$ and $q_0$ denote their values at $m = 2.6~\text{GeV}/c^2$. Here, $\mathcal{B}_j$ represents the partial wave BF of $\psi(3686)\to \gamma X_j,\ X_j\to \Lambda\bar{\Lambda}$.

The quantity $\rho^2_{j}$ corresponds to the squared modulus of the partial wave amplitude, which encodes the dynamical information of the $\Lambda\bar{\Lambda}$ system without any detector efficiency or kinematic effects. Through the normalization constraint above, the PHSP integral of $\rho^2_{j}$ directly equals to the partial decay width of the $\psi(3686)\to \gamma X_j,\ X_j\to \Lambda\bar{\Lambda}$.

\section{Appendix E: Fit Models}

The formulae used to fit the \(^1S_0\) partial wave lineshape are as follows:

\begin{itemize}
\item BW model: Sum of BW resonances
\begin{equation}
    A(m) = \sum_{k=0}^{n} \frac{c_{k}}{m_{k}^2 - m^2 - i\Gamma_{k}m_k},
\end{equation}
where $m_k$ and $\Gamma_{k}$ are the mass and width of the $k$-th resonance, respectively. $c_{k}$ is the complex coefficient.

\item $K$-matrix (KM) model~\cite{pdg2024}
\begin{equation}
\begin{aligned}
    A(m) &= \frac{P}{1-i\rho_c K },\text{~with~}K &= \sum_{k=0}^{n-1}  \frac{g_{k}^2}{m_{k}^2 - m^2}, \text{~and~}P &=  \sum_{k=0}^{n-1} \frac{\beta_{k}g_{k}}{m_{k}^2 - m^2},\\
\end{aligned}
\label{eq:K-matrix}
\end{equation}
where $\rho_c=\frac{1}{16\pi}\frac{2|q|}{\sqrt{s}}$ is a factor related to the two-body PHSP, and $|q|$ is the break-up momentum of the particles in the center-of-mass frame. $n=3$ is the number of poles. $m_{k}$ is referred to as the bare mass, $g_{k}$ represents the bare couplings and $\beta_{k}$ represents a transition driven by the coupling of the bare resonance to the source.

\end{itemize}

As an additional consistency check, we implement a third model that replaces the first BW propagator with a subthreshold pole term:
\begin{equation}
    A(m) = \frac{c_{0}}{a - m^2} + \sum_{k=1}^{n-1} \frac{c_{k}}{m_{k}^2 - m^2 - i\Gamma_{k}m_{k}},
\end{equation}
where the first term is a real-axis pole term with $a<4 m_{\Lambda}^2$. This  model yields results highly consistent with the $K$-matrix model: $m_1 = 1901 \pm 65$~MeV/$c^2$, $m_2 = 2618 \pm 16$~MeV/$c^2$, $\Gamma_2 = 93 \pm 36$~MeV.

\section{Appendix F: Extracting the Signal Pole Amplitude from the $K$-Matrix Model}

For the single-channel multi-pole $K$-matrix model, which does not explicitly describe the amplitude forms of individual poles, we employ a ``singularity removal" technique to extract single-pole amplitudes. This enables the calculation of each pole’s BFs within the K-matrix framework. The method is illustrated using a two-pole scenario and can be directly extended to cases with multiple poles.

By examining the K-matrix expression given in Eq.~\eqref{eq:K-matrix}, it can be seen that multiplying both the numerator and denominator by the factor $(m_0^2 - m^2)(m_1^2 - m^2)$ allows the formula to be rewritten as follows:
\begin{equation}
\begin{aligned}
    A(s) &= \frac{\beta_{0}g_{0}(m_1^2 - s) + \beta_{1}g_{1}(m_0^2 - s) }{(m_0^2 - s)(m_1^2 - s) - i\rho_c (g_{0}^2(m_1^2 - s) + g_{1}^2(m_0^2 - s))} = \frac{N(s)}{D(s)},
\end{aligned}
\end{equation}
where we change the independent variable from $m$ to $s=m^2$ for convenience.

In general, the denominator \( D(s) \) has two zeros, \( s_0 \) and \( s_1 \), in the complex \( s \)-plane, obtainable through numerical methods. The numerator term \( N(s) \), a linear polynomial in \( s \) for the two-pole case, can be expressed as
\begin{equation}
    N(s) =  c_0 (s_1 - s) + c_1 (s_0 - s),
\end{equation}
where $c_0$ and $c_1$ are the coefficients to be solved by
\begin{equation}
 c_0 (s_1 - s) + c_1 (s_0 - s) = \beta_{0}g_{0}(m_1^2 - s) + \beta_{1}g_{1}(m_0^2 - s).
\end{equation}
Then  the K-matrix amplitude $A(s)$ can be rewritten as:
\begin{equation}
    A(s) = \frac{c_0 (s_1 - s)}{D(s)} + \frac{c_1 (s_0 - s)}{D(s)}  = A_0(s) + A_1(s).
\end{equation}
Recall that \( s_0 \) and \( s_1 \) are the two zeros of \( D(s) \). Consequently, \( A_0(s) \) has a pole only at \( s_0 \), while \( A_1(s) \) has a pole only at \( s_1 \). To apply the \lq\lq singularity removal" method - by using the solved pole parameters to decompose the numerator polynomial, we can eliminate the singularity caused by other poles and isolate the amplitude of each pole.

\section{Appendix G: BF Calculation for the $^1S_0$ Partial Wave}

The fit fraction corresponding to the $k$-th pole is given by

\begin{equation}
    FF_k = \frac{\int |A_k(m)|^2 P(m) dm}{\int |A(m)|^2 P(m) dm},
\end{equation}
where $P(m) = p q $ is PHSP factor for the sequential decay $\psi(3686) \to \gamma X$, $X \to \Lambda\bar{\Lambda}$. $p$ and $q$ are the breakup momenta of  $X$ and $\Lambda$ in the center-of-mass frame of their mother particles, respectively.

The BF of each $^1S_0$ resonance can be determined as
\begin{equation}
     \mathcal{B}_k = FF_k  \mathcal{B}(^1S_0).
\end{equation}

\begin{table*}
\renewcommand{\arraystretch}{1.2}
\fontsize{9pt}{11pt}\selectfont
\centering
\setlength{\tabcolsep}{10pt}
\caption{Detailed information for each bin. Here, $m_{i}$ and $w_{i}$ denote the center mass and bin width, respectively. $N_{\mathrm{data}}$, $N_{\mathrm{bg}}$, and $N_{\mathrm{signal}}$ represent the number of data, background, and signal events, respectively. $\epsilon_{i}$ is the efficiency. For the $\rho^2$ values, the first uncertainty is statistical and the second is systematic.}
\label{tab:amp_square}
\begin{tabular}{cccccccc}
\hline
\hline
Bin index & \makecell{Bin range \\ (GeV/$c^2$)} & $N_{\mathrm{data}}$ & $N_{\mathrm{bg}}$ & $N_{\mathrm{signal}}$ & $\epsilon_{i}$ (\%) & \makecell{$\rho^2(^1S_0)$ \\ (eV)/(GeV/$c^2$) } & \makecell{$\rho^2(^3P_0)$ \\ (eV)/(GeV/$c^2$) } \\
\hline
  0 & (2.231, 2.256) & 127 & 2 & 125 & 13.3 & $24.39 \pm 7.98 \pm 2.23$ & $21.51 \pm 8.48 \pm 6.62$ \\
  1 & (2.256, 2.275) & 128 & 1 & 127 & 12.9 & $19.48 \pm 6.40 \pm 3.36$ & $16.64 \pm 6.68 \pm 3.20$ \\
  2 & (2.275, 2.291) & 127 & 3 & 124 & 12.5 & $15.16 \pm 6.75 \pm 1.85$ & $21.57 \pm 7.33 \pm 1.29$ \\
  3 & (2.291, 2.305) & 127 & 1 & 126 & 12.1 & $26.23 \pm 6.66 \pm 0.56$ & $12.13 \pm 6.61 \pm 0.86$ \\
  4 & (2.305, 2.321) & 127 & 1 & 126 & 11.8 & $12.92 \pm 4.95 \pm 1.64$ & $18.21 \pm 5.47 \pm 1.39$ \\
  5 & (2.321, 2.336) & 128 & 1 & 127 & 11.5 & $11.55 \pm 5.44 \pm 2.54$ & $19.80 \pm 6.04 \pm 3.06$ \\
  6 & (2.336, 2.349) & 127 & 2 & 125 & 11.2 & $16.78 \pm 6.75 \pm 5.09$ & $19.60 \pm 7.10 \pm 1.99$ \\
  7 & (2.349, 2.367) & 127 & 2 & 125 & 11.0 & $11.69 \pm 4.35 \pm 1.12$ & $11.98 \pm 4.57 \pm 0.35$ \\
  8 & (2.367, 2.385) & 128 & 3 & 125 & 11.2 & $15.67 \pm 3.92 \pm 3.13$ & $6.51 \pm 3.88 \pm 6.49$ \\
  9 & (2.385, 2.404) & 127 & 3 & 124 & 11.1 & $15.94 \pm 3.60 \pm 2.42$ & $4.35 \pm 3.33 \pm 4.52$ \\
 10 & (2.404, 2.429) & 127 & 6 & 122 & 11.4 & $7.88 \pm 2.48 \pm 1.90$ & $5.42 \pm 2.47 \pm 1.97$ \\
 11 & (2.429, 2.459) & 128 & 4 & 124 & 11.5 & $9.09 \pm 2.02 \pm 1.23$ & $2.22 \pm 1.91 \pm 2.90$ \\
 12 & (2.459, 2.488) & 127 & 6 & 121 & 12.1 & $5.70 \pm 1.76 \pm 0.21$ & $4.05 \pm 1.77 \pm 0.71$ \\
 13 & (2.488, 2.525) & 127 & 7 & 120 & 12.8 & $5.95 \pm 1.27 \pm 3.52$ & $1.10 \pm 1.16 \pm 1.81$ \\
 14 & (2.525, 2.562) & 128 & 6 & 122 & 13.5 & $4.94 \pm 1.21 \pm 0.18$ & $1.72 \pm 1.14 \pm 0.25$ \\
 15 & (2.562, 2.599) & 127 & 6 & 121 & 14.4 & $3.07 \pm 1.06 \pm 0.22$ & $2.82 \pm 1.09 \pm 0.27$ \\
 16 & (2.599, 2.630) & 127 & 5 & 122 & 15.0 & $4.43 \pm 1.10 \pm 0.52$ & $2.26 \pm 1.08 \pm 2.05$ \\
 17 & (2.630, 2.658) & 127 & 4 & 123 & 15.3 & $6.63 \pm 1.26 \pm 1.32$ & $0.55 \pm 1.23 \pm 0.59$ \\
 18 & (2.658, 2.680) & 128 & 5 & 123 & 15.9 & $7.78 \pm 1.56 \pm 0.88$ & $0.66 \pm 1.53 \pm 0.76$ \\
 19 & (2.680, 2.705) & 127 & 5 & 122 & 16.1 & $5.08 \pm 1.28 \pm 0.50$ & $2.54 \pm 1.25 \pm 2.56$ \\
 20 & (2.705, 2.727) & 127 & 2 & 125 & 16.4 & $6.51 \pm 1.31 \pm 0.64$ & $1.57 \pm 1.20 \pm 0.58$ \\
 21 & (2.727, 2.753) & 128 & 4 & 124 & 16.8 & $5.90 \pm 1.19 \pm 0.65$ & $1.16 \pm 1.08 \pm 1.19$ \\
 22 & (2.753, 2.775) & 127 & 3 & 124 & 17.1 & $7.98 \pm 0.79 \pm 0.96$ & $0.00 \pm 0.44 \pm 0.01$ \\
 23 & (2.775, 2.798) & 127 & 6 & 121 & 17.2 & $4.95 \pm 1.23 \pm 0.37$ & $2.20 \pm 1.16 \pm 2.22$ \\
 24 & (2.798, 2.823) & 128 & 2 & 126 & 17.5 & $5.90 \pm 1.21 \pm 0.25$ & $1.11 \pm 1.08 \pm 1.12$ \\
 25 & (2.823, 2.849) & 127 & 5 & 122 & 17.5 & $6.54 \pm 0.66 \pm 1.05$ & $0.00 \pm 1.43 \pm 0.03$ \\
 26 & (2.849, 2.874) & 127 & 4 & 123 & 17.8 & $5.08 \pm 1.15 \pm 0.27$ & $1.57 \pm 1.06 \pm 0.67$ \\
 27 & (2.874, 2.899) & 127 & 7 & 120 & 17.8 & $6.03 \pm 1.19 \pm 1.03$ & $0.50 \pm 1.18 \pm 0.76$ \\
 28 & (2.899, 2.921) & 128 & 3 & 125 & 18.1 & $7.69 \pm 1.42 \pm 1.21$ & $0.19 \pm 2.38 \pm 0.74$ \\
 29 & (2.921, 2.940) & 126 & 1 & 125 & 18.3 & $7.29 \pm 1.41 \pm 0.62$ & $1.54 \pm 1.24 \pm 1.63$ \\
 30 & (2.940, 2.952) & 120 & 1 & 119 & 18.3 & $10.30 \pm 2.65 \pm 1.13$ & $3.54 \pm 2.48 \pm 0.83$ \\
 31 & (2.952, 2.963) & 140 & 1 & 139 & 18.6 & $15.43 \pm 2.60 \pm 1.73$ & $0.53 \pm 3.35 \pm 0.80$ \\
 32 & (2.963, 2.975) & 234 & 1 & 233 & 18.5 & $24.63 \pm 3.36 \pm 2.36$ & $2.14 \pm 2.97 \pm 2.70$ \\
 33 & (2.975, 2.987) & 267 & 1 & 266 & 17.7 & $29.33 \pm 3.62 \pm 1.15$ & $1.38 \pm 3.41 \pm 1.96$ \\
 34 & (2.987, 2.999) & 166 & 0 & 166 & 18.7 & $19.23 \pm 1.67 \pm 0.91$ & $0.00 \pm 0.97 \pm 0.01$ \\
 35 & (2.999, 3.011) & 57 & 1 & 56 & 19.8 & $6.46 \pm 0.91 \pm 0.68$ & $0.00 \pm 0.53 \pm 0.02$ \\
 36 & (3.011, 3.022) & 28 & 0 & 28 & 19.7 & $3.03 \pm 1.14 \pm 0.19$ & $0.21 \pm 1.66 \pm 0.22$ \\
 37 & (3.022, 3.034) & 17 & 0 & 17 & 19.4 & $1.42 \pm 0.85 \pm 0.49$ & $0.51 \pm 0.84 \pm 0.33$ \\
 38 & (3.034, 3.046) & 17 & 0 & 17 & 19.2 & $0.00 \pm 0.60 \pm 0.01$ & $2.00 \pm 0.54 \pm 0.36$ \\
 39 & (3.046, 3.081) & 32 & 1 & 31 & 18.8 & $0.00 \pm 0.28 \pm 0.01$ & $1.54 \pm 0.32 \pm 0.31$ \\
\hline
\hline
\end{tabular}
\end{table*}

\section{Appendix H: Model-Dependent PWA}

\begin{figure}[H]
    \centering
    \subfigure[$M_{\Lambda\bar{\Lambda}}$ distribution]
    {
        \includegraphics[width=0.31\textwidth]{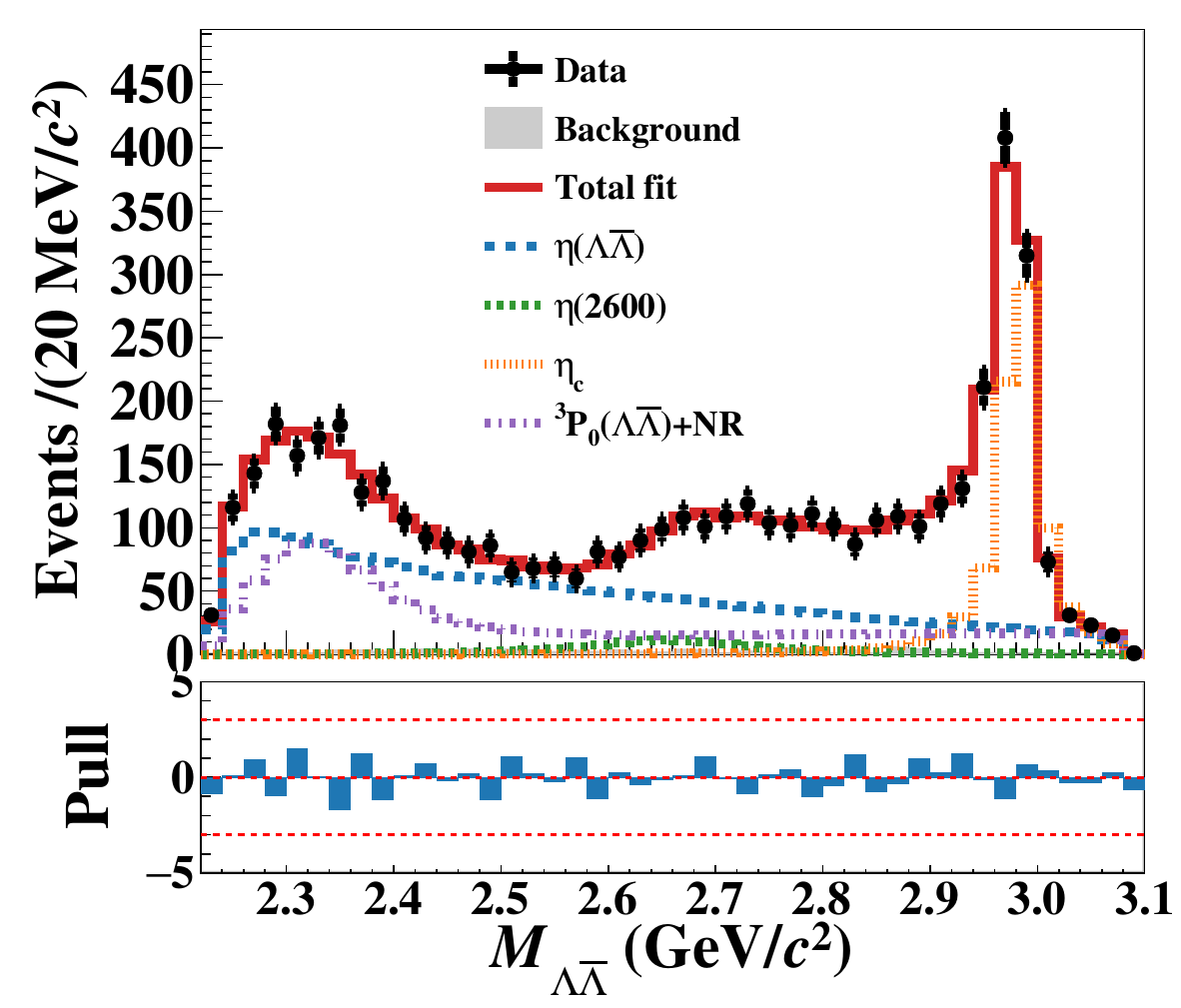}
    }
    \subfigure[$M_{\gamma\Lambda}$ distribution]
    {
        \includegraphics[width=0.31\textwidth]{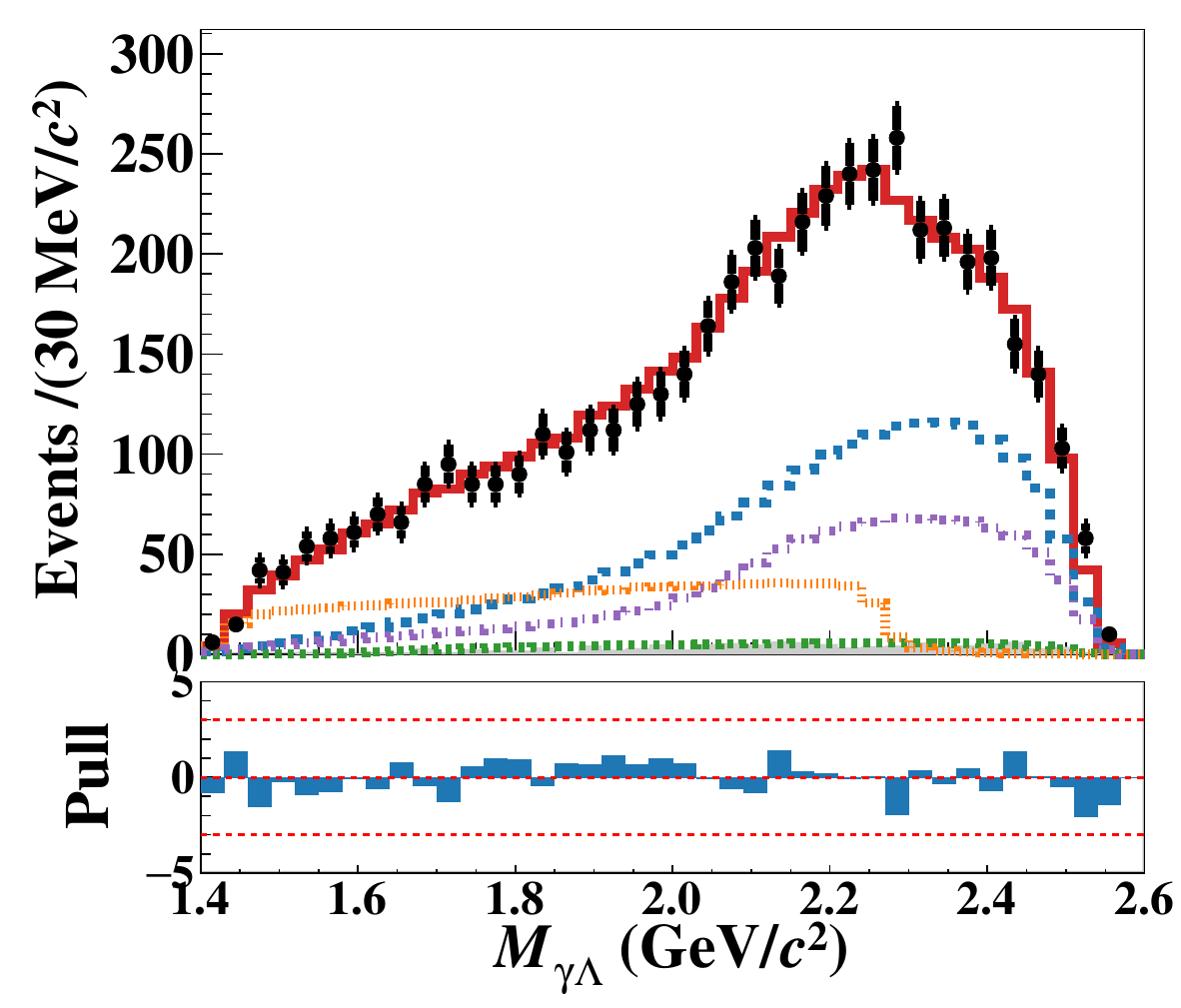}
    }
    \subfigure[$M_{\gamma\bar{\Lambda}}$ distribution]
    {
        \includegraphics[width=0.31\textwidth]{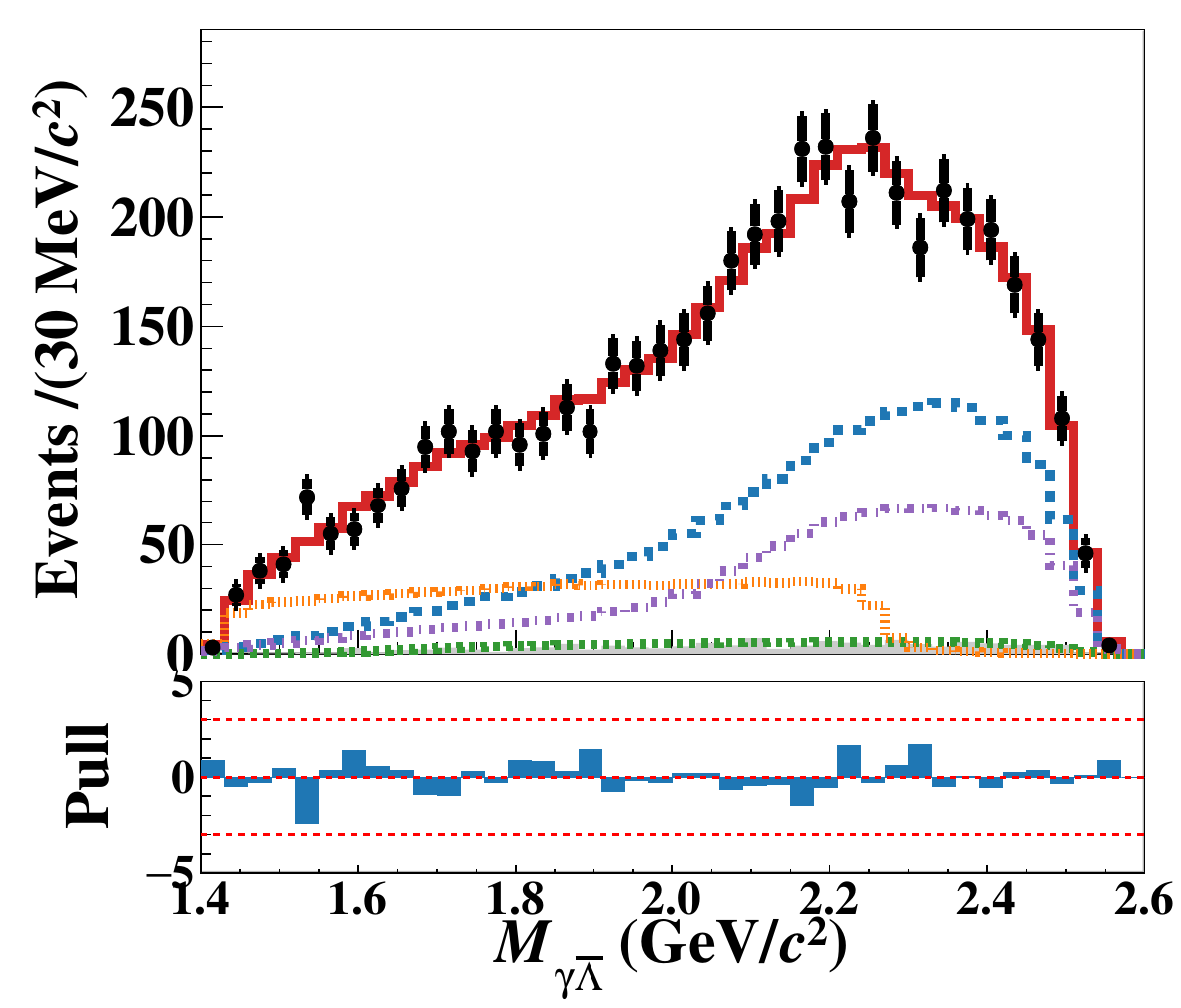}
    }
    \caption{Model-dependent fit results showing (a) the $M_{\Lambda\bar{\Lambda}}$ invariant mass distribution, (b) the $M_{\gamma\Lambda}$ distribution, and (c) the $M_{\gamma\bar{\Lambda}}$ distribution. The black circles with error bars represent background-subtracted data, the red histogram shows the total fit, and the colored histograms show individual resonance contributions: $\eta(\Lambda\bar{\Lambda})$  (blue), $\eta(2600)$ (green), $\eta_c$ (orange), $^3P_0(\Lambda\bar{\Lambda})$  (purple).}
    \label{fig:md_fit_result_all}
\end{figure}

\begin{figure}[H]
    \centering
    \subfigure[$^1S_0$ partial wave]
    {
        \includegraphics[width=0.45\textwidth]{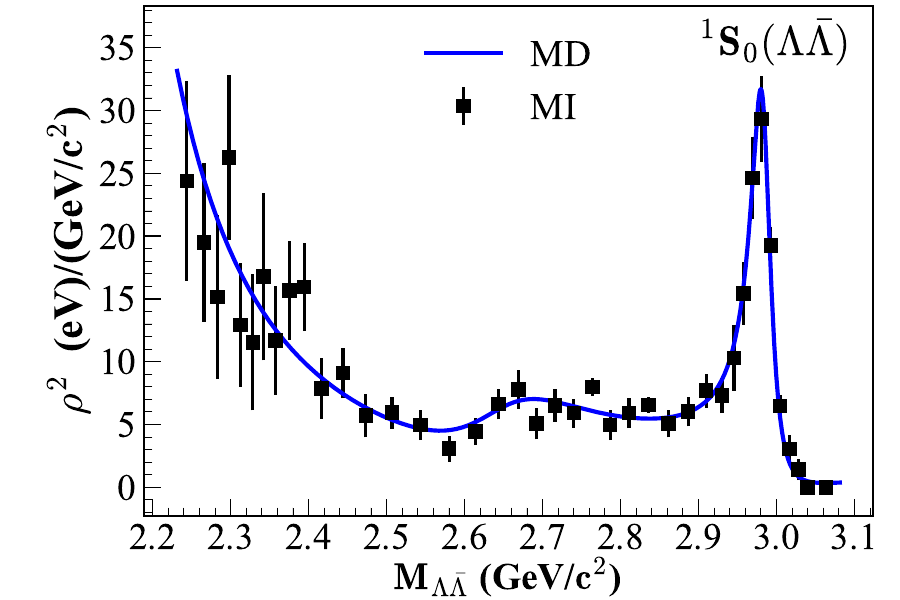}
    }
    \subfigure[$^3P_0$ partial wave]
    {
        \includegraphics[width=0.45\textwidth]{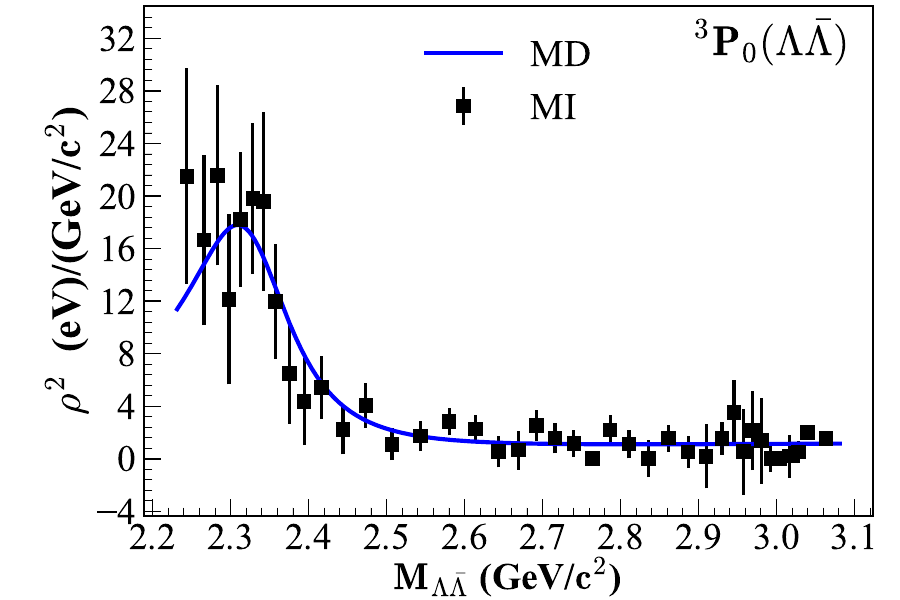}
    }
    \caption{Comparison between MD and MI results for (a) the $^1S_0$ partial wave and (b) the $^3P_0$ partial wave. The black squares with error bars represent MI measurements with statistical uncertainties only, while the solid curves show the MD results.}
    \label{fig:md_mi_comparison}
\end{figure}

As a supplement to the model-independent (MI) PWA presented in the main text, we perform a model-dependent (MD) analysis to provide an alternative description of the invariant mass distributions and to cross-check the MI results. The MD parametrization employs a subthreshold pole model for the $^1S_0$ threshold enhancement structure $\eta(\Lambda\bar{\Lambda})$, combined with BW propagators for the $\eta(2600)$ and $\eta_c$ resonances, with their masses and widths as free parameters. For the $^3P_0$ partial wave, we use a single BW resonance plus a non-resonant component. The model parameters are determined through an unbinned maximum likelihood fit to the full dataset. Figure~\ref{fig:md_fit_result_all} shows the MD fit results for the $M_{\Lambda\bar{\Lambda}}$, $M_{\gamma\Lambda}$, and $M_{\gamma\bar{\Lambda}}$ invariant mass distributions.  Figure~\ref{fig:md_mi_comparison} presents a comparison between the MD results and the MI amplitude measurements  for both partial waves.  The MI measurements are in good agreement with the MD results within statistical uncertainties.

To assess the statistical significance of the $\eta(2600)$ resonance within the MD framework, we perform a likelihood ratio test comparing two nested hypotheses: the nominal model including the $\eta(2600)$ component and an alternative model excluding it.  The nominal fit yields $\Delta\mathrm{NLL} = 30.8$ with $\Delta n = 4$ degrees of freedom, corresponding to a statistical significance of $7.0\sigma$.

\end{appendices}

\end{document}